%% file: CNSAFInalJournal.tex
\documentclass[a4paper,11pt]{witpress}
 
\DeclareRobustCommand*\cal{\@fontswitch\relax\mathcal}

\usepackage[utf8]{inputenc}

\usepackage[T1]{fontenc}
\usepackage[normalem]{ulem}
 \usepackage[english,french]{babel}
\usepackage{verbatim}
\usepackage{graphicx}
\usepackage{latexsym}
\usepackage{graphics}
\usepackage{epsfig}
\usepackage{amsmath}
\usepackage{rotating}
\usepackage{url}
\usepackage{fancyhdr}
\usepackage{tabularx}
\usepackage{array}
\usepackage{enumitem}
\usepackage{amsthm}
\usepackage{amsmath}
\usepackage{amssymb}
\usepackage{makeidx}
\usepackage{multirow}

\usepackage{multicol}

\usepackage{float}
\usepackage{geometry}
\usepackage{tikz}
\usepackage{pstricks}
\usetikzlibrary{trees}
 \usepackage{cite}
\usepackage{url}
 \usepackage{times}
\usepackage{array}
\date{}

\theoremstyle{definition}
\newtheorem{thm}{Theorem}[section]

\newtheorem{lem}[thm]{Lemma}
\newtheorem{pre}[thm]{Proposition}

\newtheorem{defini}[thm]{Definition}
\newtheorem{exemple}[thm]{Example}

\newtheorem{preuve}{Proof}
\newtheorem*{preuve2}{Proof}

\newtheorem*{remarque}{Remark}

\begin{document}
{\renewcommand{\arraystretch}{1}%

   \selectlanguage{english}
\setlength{\parskip}{0em}
\setlength{\parindent}{0em}
\title{\sc{A Semi-Decidable Procedure to Analyze Cryptographic Protocols for Secrecy}}

\begin{center}
       {\LARGE \sc \textbf{A Semi-Decidable Procedure for Secrecy in}}     \\ 
          {\LARGE \sc \textbf{}} \\
       {\LARGE \sc \textbf{Cryptographic Protocols }} 
    \vspace{1cm}

         { \Large{Jaouhar Fattahi}$^{1}$ and }     {\Large{Mohamed Mejri}$^{1}$ and }         { \Large{Hanane Houmani}$^{2}$} \\


    \vspace{1em}
$^{1}${\large{LSI Group, Laval University, Quebec, Canada}}\\
$^{2}${\large{University Hassan II, Morocco}}

    \vspace{1em}
        {\large \textbf{}} \\ 
\end{center}


\section*{\large{\textit{A\normalsize{BSTRACT}}}}
\noindent \small \textit{In this paper, we present a new semi-decidable procedure to analyze cryptographic protocols for  secrecy based on a new class of functions that we call: the Witness-Functions. A Witness-Function is a reliable function that guarantees the secrecy in any protocol proved increasing once analyzed by it. Hence, the problem of  correctness becomes a problem of protocol growth. A Witness-Function operates on derivative messages in a role-based specification and introduces new derivation techniques. We give here the technical aspects of the Witness-Functions and we show how to use them in a semi-decidable procedure. Then, we analyze a variation of the Needham-Schroeder protocol and we show that a Witness-Function can also help to teach about flaws. Finally, we analyze the NSL protocol and we prove that it is correct with respect to  secrecy.}
\section*{\large{\textit{K\normalsize{EYWORDS}}}}
\noindent\small \textit{Cryptographic Protocols, Decidability, Role-based specification, Secrecy}

\normalsize

\setlength{\parskip}{0em}
\setlength{\parindent}{0em}
\input{motivation}

\input{notations}
\input{Confpreuve}
\input{ConfFonctionsInterEKEN}
\input{ConfTemoinBorneeEnEK}

\normalsize
\input{AnalyseNS}
\input{Exemple2Leger}

\input{ConfconclusionEK}
\bibliographystyle{ieeetr}
\small{
\bibliography{Ma_these} 
}
\normalsize

\end{document}

%% file: motivation.tex
\section{\sc\textbf{Introduction}}
\noindent In this paper, we present a new semi-decidable procedure for analyzing cryptographic protocols statically for the property of secrecy in a role-based specification. The main idea of this procedure is to prove the secrecy of a protocol by proving that it is increasing. Intuitively, an increasing protocol preserves secret. That means if the level of security of all atomic messages exchanged in the protocol does not decay between all receiving and sending steps in the protocol, the secret is preserved. For that, we need reliable  metrics to estimate the level of security of atomic messages. This way of seeing secrecy in protocols has been adopted in some prior works. For instance, in ~\cite{Schneider4}, Steve Schneider suggested the rank-functions to analyze protocols in CSP~\cite{Schneider96,SchneiderD04}. These functions were efficient for analyzing several protocols such the Needham-Schroeder protocol. However, using these functions dictates the protocol implementation in the CSP algebra. Besides, building these functions is not easy and their existence is not always possible~\cite{Heather}. In~\cite{Houmani1,Houmani3,Houmani8,Houmani5}, Houmani et  al.  presented universal functions called interpretation functions to statically analyze a protocol for secrecy. An interpretation function needs to meet some conditions to be "good enough" to run an analysis.  They were successful to analyze many protocols. However, we note that the conditions on these functions were very restrictive. That's why only two  functions had been given: DEK and DEKAN. Naturally, less we have restrictions on functions, more we have chance to define many of them and therefore to prove the correctness of a larger range of protocols. In fact, one function may fail to prove the growth of a protocol but another may do so. In this regard, we think that the condition of full-invariance by substitution in Houmani's  wrok is the most blocking one. This condition is though very important since it enables any decision made on messages of the generalized roles (messages with variables) to be propagated to valid traces (closed messages). Since the goal of our approach is to build as many functions as we can, we believe that if we liberate a function from this condition, we will be able to build several functions. However, liberating a function from a condition may oblige us to take extra precautions when using it. In this paper, we present the Witness-Functions as new metrics to analyze cryptographic protocols. A Witness-Function is tightly linked to an interpretation function but does not need the full-invariant by substitution property. In fact, a Witness-Function provides two attractive bounds that are independent of substitution. This fully replaces any need to this property. We also introduce the notion of derivative messages by using new derivation techniques. We exhibit the theorem of protocol analysis with the Witness-Functions. This theorem defines a semi-decidable procedure for analyzing cryptographic protocols. Finally, we run an analysis on two protocols. First, we run an analysis on a variation of Needham-Schroeder protocol in which we show that a Witness-Function could even teach about flaws. Then, we run an analysis on NSL protocol where we prove that it is correct with respect to secrecy.

%% file: notations.tex
\section{\sc\textbf{Preliminary and Notations}}
\noindent Here, we give some conventions and notations that we use in this paper. 
\begin{itemize}
\item[+] We denote by  ${\mathcal{C}}=\langle{\mathcal{M}},\xi,\models,{\mathcal{K}},{\mathcal{L}}^\sqsupseteq,\ulcorner.\urcorner\rangle$ the context of verification in which our analysis is run. It contains the parameters that affect the analysis of a protocol:
\begin{itemize}
\item[$\bullet$] ${\mathcal{M}}$ : is a set of messages built from the signature $\langle\mathcal{N}$,$\Sigma\rangle$ where ${\mathcal{N}}$ is a set of atomic names (nonces, keys, principals, etc.) and $\Sigma$ is a set of functions ($enc$:\!: encryption\index{Encryption}, $dec$:\!: decryption\index{Décryption}, $pair$:\!: concatenation (that we denote by "." here), etc.). i.e. ${\mathcal{M}}=T_{\langle{\mathcal{N}},\Sigma\rangle}({\mathcal{X}})$.  We denote by $\Gamma$ the set of  substitutions from $ {\mathcal{X}} \rightarrow {\mathcal{M}}$.
We denote by $\mathcal{A}$ all the atomic messages in ${\mathcal{M}},$  by ${\mathcal{A}}(m)$ the set of atomic messages (or atoms) in $m$ and by ${\mathcal{I}}$ the set of principals including the intruder $I$. We denote by $k{^{-1}}$ the reverse form of a key $k$ and we assume that $({k^{-1}})^{-1}=k$.
\item[$\bullet$] $\xi$ : is the equational theory in which the algebraic properties of the functions in $\Sigma$ are described by equations. e.g. $dec(enc(x,y),y^{-1})=x$. 
\item[$\bullet$] $\models_{\mathcal{C}}$ : is the inference system of the intruder under the equational theory. Let $M$ be a set of messages and $m$ a message. $M$ $\models_{\mathcal{C}}$ $m$ means that the intruder is able to infer $m$ from $M$ using her capacity. We extend this notation to traces as follows: $\rho$ $\models_{\mathcal{C}}$ $m$ means that the intruder can  infer $m$ from the messages exchanged in the trace $\rho$.
We suppose that the intruder has the full control of the net as described by Dolev-Yao model in~\cite{DolevY1}. That is to say that she can intercept, delete, redirect and modify messages. She knows the public keys of all agents. She knows her private keys and the keys that she shares with other agents. She can encrypt or decrypt any message with known keys. Generically, the intruder has the following rules of building messages:\\

\begin{center}
\begin{itemize}
  \item[] $(int):$ $\frac{\square}{M\models_{\mathcal{C}} m}[m\in M\cup K(I)] $\\
$ $\\
  \item[] $(op):$$\frac{M\models_{\mathcal{C}} m_1,..., M\models_{\mathcal{C}} m_n }{M\models_{\mathcal{C}} f(m_1,...,m_n)}[f\in \Sigma]$\\
$ $\\
  \item[] $(eq):$$\frac{M\models_{\mathcal{C}} m', m'=_{\mathcal{C}} m}{M \models_{\mathcal{C}} m}$, with  $(m'=_{\mathcal{C}} m)\equiv (m'=_{\xi_{({\mathcal{C}})}} m)$\\
$ $\\
\end{itemize}
\end{center}

\begin{exemple}
$ $\\

The intruder capacity can be described by the following rules:\\
\begin{itemize}
  \item[] $(int):$ $\frac{\square}{M\models_{\mathcal{C}} m}[m\in M\cup K(I)] $\\

  \item[] $(concat):$$\frac{M\models_{\mathcal{C}} m_1,M\models_{\mathcal{C}} m_2 }{M\models_{\mathcal{C}} m_1.m_2}$\\

  \item[] $(deconcat):$$\frac{M\models_{\mathcal{C}} m_1.m_2}{M\models_{\mathcal{C}} m_i}[i\in \{1,2\}]$\\

  \item[] $(dec):$$\frac{M\models_{\mathcal{C}} k,M\models_{\mathcal{C}} m_k }{M\models_{\mathcal{C}} m}$\\

  \item[] $(enc):$$\frac{M\models_{\mathcal{C}} k,M\models_{\mathcal{C}} m }{M\models_{\mathcal{C}} \{m\}_k}$\\

\end{itemize}
\end{exemple}

In this example, from a set of messages, an intruder can infer any message in this set. She can encrypt any message when she holds the encryption key. She can  decrypt any message when she holds the decryption key and concatenate any two messages and deconcatenate them.

\item[$\bullet$] ${\mathcal{K}}$ : is a function from ${\mathcal{I}}$ to ${\mathcal{M}}$, that returns to any agent a set of atomic messages describing her initial knowledge. We denote by $K_{{\mathcal{C}}}(I)$ the initial knowledge of the intruder, or simply $K(I)$ where the context is obvious.
\item[$\bullet$] ${\mathcal{L}}^\sqsupseteq$ : is the  lattice of security $({\mathcal{L}},\sqsupseteq, \sqcup, \sqcap, \bot,\top)$ used to assign security levels to  messages. 
An example of a lattice is $ (2^{\mathcal{I}},\subseteq,\cap,\cup,\mathcal{I}, \emptyset)$ that will be used to attribute to an atomic message $\alpha$ the set of agents that are authorized to know it. 
\item[$\bullet$] $\ulcorner .\urcorner$ : is a partial function that attributes a value of security (or type) to a message in ${\mathcal{M}}$. Let $M$ be a set of messages and $m$ a message. We write $\ulcorner M \urcorner \sqsupseteq \ulcorner m \urcorner$ if
$\exists m' \in M. \ulcorner m' \urcorner \sqsupseteq \ulcorner m \urcorner$
\end{itemize}
\item[+]
Our analysis is performed in a role-based specification. A role-based specification is a set of generalized roles. A generalized  role is  an   abstraction  of the protocol where  the emphasis is put on a specific agent and
where all the unknown messages, and on which the agent cannot carry out any verification, are substituted by variables. An  exponent  $i$  (the session  identifier)  is  added  to a fresh message to say that these components
change values from one run to another. A generalized role interprets how a
particular agent understands the exchanged messages. We extract it from a protocol as follows:\\
    \begin{itemize}
    \item we extract the  roles from the protocol.
    \item we substitute the unknown messages by fresh variables for each role.\\
    \end{itemize}
The roles  are extracted as follows:\\
\begin{itemize}
\item For each agent, we extract from the protocol  all  the  steps  in  which  this  principal participates.  Then, we add to this abstraction a session identifier $i$ in the steps identifiers and in the fresh values. 
\item We introduce an intruder  $I$  to express the fact that the  received messages and the sent messages are probably sent or received by the intruder.
\item Finally, we extract all  prefixes from those roles where a prefix ends by a sending step. \\

\end{itemize}

From the roles, we generate the generalized roles. In a generalized role, unknown messages are substituted by variables to express that the agent cannot be sure about its integrity or its origin. The role-based specification\index{Protocol!Role-Based Specification} expresses  the notion  of  valid  traces of a protocol. More details about the role-based specification could be found in~\cite{ContextWF,Debbabi11,Debbabi22,Debbabi33}. \\
\begin{exemple}\label{exNS}
$ $\\
Let us consider the Needham-Schroeder protocol given in Table~\ref{WLMV:protArt}.\\
\vspace*{-1\baselineskip}
\begin{table}[h]
    \caption{\label{WLMV:protArt} The Needham-Schroeder Protocol}
\center{
                 $  \begin{array}{|ll|}
\hline
                    p_1   =&\langle 1,A\rightarrow B: \{A.N_a\}_{k_b}  \rangle.  \\
                    & \langle 2,B\rightarrow A: \{N_a.N_b.B\}_{k_a}\rangle. \\
                    & \langle 3,A\rightarrow B: \{N_b\}_{k_b} \rangle. \\
\hline
                    \end{array}$
}

 \end{table}
\noindent The generalized roles of the agent $A$ are:\\
\begin{center}
           $\begin{array}{l}\begin{array}{lllllll}
                    {\mathcal A}_G ^1 =& \langle  i.1,&  A    & \rightarrow & I(B)&  :&  \{A.N_a^i\}_{k_b} \rangle
                    \end{array}\\
                    \\
                    \begin{array}{lllllll}
                    {\mathcal A}_G ^2=& \langle  i.1,&  A    & \rightarrow & I(B)&  :&  \{A.N_a^i\}_{k_b} \rangle .\\
                    & \langle i.2,&  I(B) & \rightarrow & A   &  :&  \{N_a^i.X.B\}_{k_a} \rangle .\\
                    & \langle i.3,&  A    & \rightarrow & I(B)&  :&  \{X\}_{k_b}\rangle
                    \end{array}\\
             \end{array}$
\end{center}
\normalsize
$ $\\
\noindent The generalized roles of the agent $B$ are:\\
\begin{center}
            $\begin{array}{l}\begin{array}{lllllll}
                {\mathcal B}_G ^1=& \langle i.1,&  I(A) & \rightarrow & B   &  :&  \{A.Y\}_{k_b}\rangle .\\
                        & \langle i.2,&  B    & \rightarrow & I(A)&  :& \{Y.N_b^i.B\}_{k_a} \rangle  \\
                    \end{array}\\
\\
                    \begin{array}{lllllll}
                {\mathcal B}_G ^2=& \langle i.1,&  I(A) & \rightarrow & B   &  :&  \{A.Y\}_{k_b}\rangle .\\
                        & \langle i.2,&  B    & \rightarrow & I(A)&  :& \{Y.N_b^i.B\}_{k_a} \rangle .\\
                        & \langle i.3,&  I(A) & \rightarrow & B   &  :& \{N_b^i\}_{k_b}\rangle\\
                    \end{array}\\
                    \end{array}$
\end{center}
\normalsize
 $ $\\
 \noindent The role-based specification  of  the  protocol  in Table~\ref{WLMV:protArt}  is  ${\mathcal
    R}_G(p_1) = \{{\mathcal A}_G ^1,~{\mathcal A}_G ^2,~{\mathcal B}_G^1,~{\mathcal B}_G ^2$\}. \\
\end{exemple}
\begin{exemple}\label{exNSL}
$ $\\
Let us consider the NSL protocol given in Table~\ref{WLMV:protArtNSL}.\\
\begin{table}[h]
    \caption{\label{WLMV:protArtNSL} The NSL Protocol}
\center{
                 $  \begin{array}{|ll|}
\hline
                    p_2   =&\langle 1,A\rightarrow B: \{N_a.A\}_{k_b} \rangle.  \\
                    & \langle 2,B\rightarrow A: \{B.N_a\}_{k_a}.\{B.N_b\}_{k_a}\rangle. \\
                    & \langle 3,A\rightarrow B: A.B.\{N_b\}_{k_b}\rangle \\
\hline
                    \end{array}$
}

 \end{table}
The generalized roles of the agent $A$ are:\\
\begin{center}
           $\begin{array}{l}\begin{array}{lllllll}
                    {\mathcal A}_G ^{'1} =& \langle  i.1,&  A    & \rightarrow & I(B)&  :&  \{N_a^i.A\}_{k_b} \rangle
                    \end{array}\\
                    \\
                    \begin{array}{lllllll}
                    {\mathcal A}_G ^{'2}=& \langle  i.1,&  A    & \rightarrow & I(B)&  :&  \{N_a^i.A\}_{k_b} \rangle .\\
                    & \langle i.2,&  I(B) & \rightarrow & A   &  :&  \{B.N_a^i\}_{k_a}.\{B.X\}_{k_a} \rangle .\\
                    & \langle i.3,&  A    & \rightarrow & I(B)&  :&  A.B.\{X\}_{k_b}\rangle
                    \end{array}\\
             \end{array}$
\end{center}
\normalsize
The generalized roles of the agent $B$ are:\\
\begin{center}
            $\begin{array}{l}\begin{array}{lllllll}
                {\mathcal B}_G ^{'1}=& \langle i.1,&  I(A) & \rightarrow & B   &  :&  \{Y.A\}_{k_b} \rangle .\\
                        & \langle i.2,&  B    & \rightarrow & I(A)&  :&  \{B.Y\}_{k_a}.\{B.N_b^i\}_{k_a} \rangle  \\
                    \end{array}\\
                    \begin{array}{lllllll}
\\
                {\mathcal B}_G ^{'2}=& \langle i.1,&  I(A) & \rightarrow & B   &  :&  \{Y.A\}_{k_b} \rangle .\\
                        & \langle i.2,&  B    & \rightarrow & I(A)&  :& \{B.Y\}_{k_a}. \{B.N_b^i\}_{k_a} \rangle .\\
                        & \langle i.3,&  I(A) & \rightarrow & B   &  :&  A.B.\{N_b^i\}_{k_b} \rangle\\
                    \end{array}\\
                    \end{array}$
\end{center}
\normalsize
 $ $\\
    The role-based specification  of  the  protocol  in Table~\ref{WLMV:protArtNSL}  is  ${\mathcal
    R}_G(p_2) = \{{\mathcal A}_G ^{'1},~{\mathcal A}_G ^{'2},~{\mathcal B}_G^{'1},~{\mathcal B}_G ^{'2}$\}. 

\end{exemple}
\item[+] A valid trace is an interleaving of substituted generalized roles where each message sent by the intruder can be generated by her using her capacity and by the received messages. We denote by $ [\![p]\!]$ the set of valid traces generated by $p$.
\item[+] We denote by ${\mathcal{M}}_p^{\mathcal{G}}$ the set of messages (with variables) in $R_G(p)$, by ${\mathcal{M}}_p$ the set of closed messages generated by substitution in ${\mathcal{M}}_p^{\mathcal{G}}$.  We denote by $R^+$ (respectively $R^-$) the set of sent messages (respectively received messages) by a honest agent in the role $R$. Conventionally, we devote the uppercase symbols for sets or sequences of elements and the lowercase for single elements. For example, $M$ denotes a set of messages, $m$ a single message, $R$ a role composed of a sequence of steps, $r$ a step and $R.r$ the role ending by the step $r$.
 \item[+] In our analysis, no restriction on the size of messages or the number of sessions in the protocols is made.
\end{itemize}


%% file: Confpreuve.tex
\section{\sc\textbf{Increasing protocols do not reveal secrets}}\label{sectionPreuveThFond}

\noindent To analyze a protocol, we need  functions to estimate the security level of every atomic message.
In this section, we give sufficient conditions on a function $F$ to guarantee that it is enough good (or reliable) to run an analysis and we show that an increasing protocol is correct with respect to the secrecy property when analyzed with such functions.

\subsection{${\mathcal{C}}$-reliable functions}

\noindent A function $F$ is said to be well-formed when it returns the lowest value  in the lattice, denoted by $\bot$, for an atomic message $\alpha$ that appears in clear. It returns for it in the union of two sets, the minimum "$\sqcap$" of the two values calculated in each set separately. It returns the uppermost value, denoted by "$\top$", if it does not appear in this set. These facts are expressed by the definition \ref{bienforme}.\\

\begin{defini}(Well-formed function)\label{bienforme}\index{function}
$ $\\
Let $F$ be a function and ${\mathcal{C}}$ a context of verification. \\
$F$ is well-formed in ${\mathcal{C}}$ if:\index{Fonction well-formed}\\
$\forall M,M_1,M_2 \subseteq {\mathcal{M}}, \forall \alpha \in {\mathcal{A}}({\mathcal{M}}) \mbox{:}\\$
\[
\left\{
    \begin{array}{lll}
         F(\alpha,\{\alpha\})&= & \bot \\
         F(\alpha, {M}_1 \cup {M}_2)&= & F(\alpha, {M}_1)\sqcap F(\alpha,{M}_2) \\
         F(\alpha,{M})& =&\top, \mbox{ if } \alpha \notin {\mathcal{A}}({M})  \\
    \end{array}
\right.
\]
\end{defini}

\noindent A function $F$ is said to be full-invariant-by-intruder if when it attributes a security level  to a message $\alpha$ in a set of messages $M$, the intruder can never produce another message $m$ that decrease this level  (i.e. $F(\alpha,m) \sqsupseteq F(\alpha,{M})$) using her capacity in the context of verification, except when $\alpha$ is intended to be known by the intruder (i.e. $\ulcorner K(I) \urcorner \sqsupseteq \ulcorner \alpha \urcorner$). This fact is expressed by the definition \ref{spi}.\\

\begin{defini}(Full-invariant-by-intruder function)\label{spi}\index{Stabilité par intrus}
$ $\\
Let $F$ be a function and ${\mathcal{C}}$ a context of verification.  \index{function}\\
$F$ is full-invariant-by-intruder in ${\mathcal{C}}$ if:\\
$
\forall {M} \subseteq {\mathcal{M}}, m\in {\mathcal{M}}. {M} \models_{\mathcal{C}} m \Rightarrow \forall \alpha \in {\mathcal{A}}(m). (F(\alpha,m) \sqsupseteq F(\alpha,{M}))  \vee (\ulcorner K(I) \urcorner \sqsupseteq \ulcorner \alpha \urcorner)\\
$
\end{defini}

\noindent A function $F$ is said to be reliable if it is well-formed and full-invariant-by-intruder. This fact is expressed by the definition \ref{relF}.\\

\begin{defini}(Reliable function)\label{relF}
$ $\\
Let $F$ be a function and ${\mathcal{C}}$ a context of verification.
\[F  \mbox { is }{\mathcal{C}}\mbox{-reliable}  \mbox{ if } F \mbox{ is well-formed and } F \mbox{ is full-invariant-by-intruder in } {\mathcal{C}}.\]
\end{defini} 


\noindent A protocol $p$ is said to be $F$-increasing when every principal generates increasingly valid traces (substituted generalized roles) that never decrease the security levels of received components. The estimation of the value of security of every atom is performed by $F$. This fact is expressed by the definition \ref{ProAbsCroiArticle}.\\


\begin{defini}($F$-increasing protocol)\label{ProAbsCroiArticle}
$ $\\
Let $F$ be a function, ${\mathcal{C}}$ a context of verification  and $p$ a protocol.\\
$p$ is $F$-increasing in ${\mathcal{C}}$ if:\\
$\forall R.r \in R_G(p),\forall \sigma \in \Gamma: {\mathcal{X}} \rightarrow {\mathcal{M}}_p  \mbox{ we have: }$
\[
\forall \alpha \in {\mathcal{A}}({\mathcal{M}}_p).F(\alpha, r^+\sigma)\sqsupseteq \ulcorner \alpha \urcorner \sqcap F(\alpha, R^-\sigma)
\]
\end{defini}

\noindent A secret disclosure consists in manipulating a valid trace of the protocol (denoted by $ [\![p]\!]$) by the intruder using her knowledge $K(I) $ in a context of verification ${{\mathcal{C}}}$, to deduce a secret $\alpha$ that she is not intended to know (expressed by: $\ulcorner K(I) \urcorner \not \sqsupseteq \ulcorner \alpha \urcorner$). This fact is expressed by the definition \ref{divulgationArticle}.\\

\begin{defini}(Secret disclosure)\label{divulgationArticle}\index{Divulgation}
$ $\\
Let $p$ be a protocol and ${\mathcal{C}}$ a context of verification.\\
We say that $p$ discloses a secret $\alpha\in {\mathcal{A}}({\mathcal{M}})$  in ${{\mathcal{C}}}$ if:
\[\exists \rho\in [\![p]\!].(\rho \models_{{\mathcal{C}}} \alpha)\wedge(\ulcorner K(I) \urcorner \not \sqsupseteq \ulcorner \alpha \urcorner)\]
\end{defini}

\begin{lem}\label{lemprincArticle}
$ $\\
Let $F$ be a ${\mathcal{C}}$-reliable function and $p$ an $F$-increasing protocol.\\
We have:
\[
\forall m \in {\mathcal{M}}. [\![p]\!] \models_{\mathcal{C}} m \Rightarrow \forall \alpha \in {\mathcal{A}}(m). (F(\alpha,m)\sqsupseteq \ulcorner \alpha \urcorner)\vee (\ulcorner K(I) \urcorner \sqsupseteq \ulcorner \alpha \urcorner)
\]

\end{lem}
\noindent\textit{See the proof 4 in~\cite{AppendixProofs}}\\

\noindent The lemma \ref{lemprincArticle} says that for any atom $\alpha$ in a message produced by an increasing protocol, its security level returned by a reliable function is kept greater or equal than its initial value in the context, if the intruder is not initially allowed to know it. Hence, initially the atom has a certain level of security. This value cannot be decreased by the intruder using her knowledge and the received messages since it is full-invariant-by-intruder. In every new step of a valid trace, involved messages are better protected since the protocol is increasing. The proof is then run by induction on the size of the trace using the reliability properties of the function in every step of the induction.\\

\begin{thm}(Theorem of Correctness of Increasing Protocols)\label{mainThArticle}
$ $\\
Let $F$ be a ${\mathcal{C}}$-reliable function and $p$ a $F$-increasing protocol.
\begin{center}
$p$ is ${\mathcal{C}}$-correct with respect to the secrecy property.
\end{center}\index{Confidentialité}

\end{thm}


\begin{preuve2}\label{PR5Article} 
$ $\\
Let's suppose that $p$ discloses an atomic secret $\alpha$. \\
\\
From the definition \ref{divulgationArticle} we have:
\begin{equation}\exists \rho\in [\![p]\!].(\rho \models_{\mathcal{C}} \alpha)\wedge(\ulcorner K(I) \urcorner \not \sqsupseteq \ulcorner \alpha \urcorner) \label{preuvthe1Article}\end{equation}
Since $F$ is a  ${\mathcal{C}}$-reliable function and $p$ an $F$-increasing protocol, we have from the lemma \ref{lemprincArticle}:
\begin{equation}(F(\alpha,\alpha)\sqsupseteq \ulcorner \alpha \urcorner)\vee (\ulcorner K(I) \urcorner \sqsupseteq \ulcorner \alpha \urcorner)\label{preuvthe2Article}\end{equation}
From \ref{preuvthe1Article} and \ref{preuvthe2Article}, we have:
\begin{equation}F(\alpha,\alpha)\sqsupseteq \ulcorner \alpha \urcorner \label{preuvthe3Article}\end{equation}
Since $F$ is well-formed in ${\mathcal{C}}$, then:
\begin{equation}F(\alpha,\alpha)= \bot \label{preuvthe4Article}\end{equation}
From \ref{preuvthe3Article} and \ref{preuvthe4Article} we have:
\begin{equation}\bot= \ulcorner \alpha \urcorner \label{preuvthe5corArticle}\end{equation}
\ref{preuvthe5corArticle} is impossible because it is contradictory with:  $\ulcorner K(I) \urcorner \not \sqsupseteq \ulcorner \alpha \urcorner$ in \ref{preuvthe1Article}.\\
\\
Then $p$ is ${\mathcal{C}}$-correct with respect to the secrecy property.\\
\end{preuve2} 

\noindent Theorem \ref{PR5Article} asserts that an increasing protocol is correct with respect to the secrecy property when analyzed with a reliable function. It is worth saying that compared to the sufficient conditions stated in~\cite{Houmani5}, we have one less. Thus, in~\cite{Houmani5}, Houmani demanded from the function an additional condition: the full-invariance by substitution. That's to say, the interpretation function has also to resist to the problem of substitution of variables. Here, we liberate our functions from this blocking condition in order to be able to build more of them. We rehouse this condition in our new definition of an increasing protocol which is required now to be increasing on valid traces (closed messages) rather than messages of the generalized roles (message with variables). Therefore, the problem of substitutions is transferred to the protocol and becomes less difficult to handle.

%% file: ConfFonctionsInterEKEN.tex
\section{\sc\textbf{Construction of reliable interpretation functions}}\label{sectionFonctionsetSelections}

\noindent As seen in the previous section, to analyze a protocol we need reliable interpretation functions to estimate the level of security of any atom in a message.  In this section, we exhibit a constructive way to build these functions. We first exhibit the way to build a generic class of reliable selections inside the protection of the most external key (or simply the external key). Then we propose specialized selections of this class. Finally we give the way to build reliable selection-based interpretation functions. Similar techniques based on selections were proposed in previous works, especially in~\cite{Houmani1,Houmani8,Houmani5} to build universal functions based on the selection of the direct key of encryption and in~\cite{Blanchet09} to check correspondences in protocols. But first of all, we present the notion of well-protected messages that have valuable properties that we will use in the definition of reliable selections.



\subsection{Protocol analysis in Well-Protected Messages}
\subsubsection{Well-Protected Messages}
Briefly, a well-protected message is a message such that every non public atom $\alpha$ in it is encrypted by at least one key $k$ such that  $\ulcorner k^{-1} \urcorner \sqsupseteq \ulcorner \alpha \urcorner$, after elimination of unnecessary keys (e.g. $e(k,d(k^{-1},m))\rightarrow m$). The main advantage of an analysis performed over a set of well-protected messages is that the intruder cannot deduce any secret when she uses only her knowledge in the context of verification (without using the protocol rules).\\

\begin{lem}{}\label{lemConservationBP}
$ $\\
Let $M$ be a set of well-protected messages in ${\mathcal{M}}$. We have:
\[
M\models_{\mathcal{C}} m \Rightarrow \forall \alpha \in {\mathcal{A}}(m).(\alpha \mbox{ is well-protected in }m) \vee (\ulcorner K(I) \urcorner \sqsupseteq \ulcorner \alpha \urcorner)
\]
\end{lem}

\noindent Lemma \ref{lemConservationBP} says that from a set of well-protected messages, all atomic messages beyond the knowledge of the intruder (i.e. $\ulcorner K(I) \urcorner \not \sqsupseteq \ulcorner \alpha \urcorner$) remain well-protected in any message that the intruder could infer. Indeed, since each atom that does not appear in clear (non-public) in this set is encrypted by at least one key $k$ such that $\ulcorner k^{-1} \urcorner \sqsupseteq \ulcorner \alpha \urcorner$,  then the intruder has to retrieve the key $k^{-1}$ before she sees $\alpha$ not well-protected in any message (clear). But, the key $k^{-1}$  is in its turn encrypted by at least one key $k'$ such that $\ulcorner k'^{-1} \urcorner \sqsupseteq \ulcorner k^{-1} \urcorner$. The proof is then conducted by induction on the encryption keys.\\

\begin{lem}(Lemma of non-disclosure of atomic secrets in well-protected messages)\label{lematomique}
\\
Let $M$ be a set of well-protected messages in ${\mathcal{M}}$ and $\alpha$ an atomic message in $M$. \\
We have:
\[M\models_{\mathcal{C}} \alpha \Rightarrow \ulcorner K(I) \urcorner \sqsupseteq \ulcorner \alpha \urcorner\]
\end{lem}


\begin{preuve2}\label{PR23Article}
$ $\\
\\
From Lemma \ref{lemConservationBP}, we have $M\models_{\mathcal{C}} \alpha$ then:
\begin{equation}(\alpha \mbox{ is well-protected in }\alpha) \vee (\ulcorner K(I) \urcorner \sqsupseteq \ulcorner \alpha \urcorner)\label{preuveCons1Article}\end{equation}
But $\alpha \mbox{ is not well-protected in }\alpha$, then we have:
\begin{equation}\ulcorner K(I) \urcorner \sqsupseteq \ulcorner \alpha \urcorner\label{preuveCons2Article}\end{equation}
\end{preuve2}
\subsection{Discussion and Assumption}

\noindent Lemma \ref{lematomique} expresses an important result. It states that from a set of well-protected messages the intruder cannot deduce any secret that she is not supposed to know when she uses only her knowledge in the context of verification (without using the protocol rules).  It is worth saying that verifying whether a protocol operates over a space of well-protected messages or not is an easy task and most of real protocols respect this condition. 


\subsection{Building reliable selections}

\noindent Now, we will focus on building selections such that when they are composed to suitable homomorphisms, provide reliable interpretation functions. The definition \ref{selecBF}  introduces the notion of a well-formed selection and the definition \ref{DefFnInterIsotrop}  introduces the notion of a full-invariant-by-intruder selection.\\

\begin{defini}(Well-formed selection)\label{selecBF}
$ $\\
Let $M,M_1,M_2\subseteq {\mathcal{M}}$ such that $M, M_1$ and $M_2$ are well-protected.\\
Let $S: {\mathcal{A}}\times {\mathcal{M}}\longmapsto{2^{\mathcal{A}}}$ be a selection.\\
We say that $S$ is well-formed in ${\mathcal{C}}$ if:
\[
\left\{
    \begin{array}{lll}
         S(\alpha,\{\alpha\})&=&{\mathcal{A}},\\
         S(\alpha, {M}_1 \cup {M}_2)&=&S(\alpha, {M}_1)\cup S(\alpha,{M}_2),\\
         S(\alpha,{M})&=&\emptyset  \mbox{, if } \alpha \notin {\mathcal{A}}({M})\\
    \end{array}
\right.
\]
\end{defini}

\noindent For an atom $\alpha$ in a set of messages $M$, a well-formed selection returns all the atoms in ${\mathcal{M}}$ if $M=\{\alpha\}$. It returns for it in the union of two sets of messages, the union of the two selections performed in each set separately. It returns the empty set if the atom does not appear in $M$.\\
\normalsize

\begin{defini}(Full-invariant-by-intruder selection)\label{DefFnInterIsotrop}
$ $\\
Let $M\subseteq {\mathcal{M}}$ such that $M$ is well-protected.\\
Let $S: {\mathcal{A}}\times {\mathcal{M}}\longmapsto{2^{\mathcal{A}}}$ be a selection.\\
We say that $S$ is full-invariant-by-intruder in ${\mathcal{C}}$ if:\\
$\forall M \subseteq {\mathcal{M}}, m \in {\mathcal{M}},$ we have:
\[
M \models_{\mathcal{C}} m\Rightarrow  \forall \alpha \in {\mathcal{A}}(m).(S(\alpha,m) \subseteq S(\alpha,M))\vee (\ulcorner K(I) \urcorner \sqsupseteq \ulcorner \alpha \urcorner)
\]

\end{defini}

\noindent The aim of a full-invariant-by-intruder selection is to provide a full-invariant-by-intruder function when composed to an adequate homomorphism that transforms its returned atoms into security levels. Since a full-invariant-by-intruder function is requested to resist to any attempt of the intruder to generate a message $m$ from any set of messages $M$ in which the level of security of an atom, that she is not allowed to know, decreases compared to it its value in $M$, a full-invariant-by-intruder selection is requested to resist to any attempt of the intruder to generate a message $m$ from any set of messages $M$ in which the selection associated to an atom, that she is not allowed to know, could be enlarged compared to the selection associated to this atom in $M$. This fact is described by the definition \ref{DefFnInterIsotrop}.\\

\begin{defini}(Reliable selection)\label{selecreliable}
$ $\\
Let $S: {\mathcal{A}}\times {\mathcal{M}}\longmapsto{2^{\mathcal{A}}}$ be a selection and ${\mathcal{C}}$ be a context of verification.
\[S  \mbox { is }{\mathcal{C}}\mbox{-reliable } \mbox{if}  \mbox{ } S \mbox{ is well-formed }  \mbox{and }  S \mbox{ is full-invariant-by-intruder in } {\mathcal{C}}.\]
\end{defini}

\subsubsection{Reliable selections inside the protection of an external key}
$ $\\
Here, we define a generic class of selections that we denote by  $S_{Gen}^{EK}$ and we prove that any instance of it is reliable under some condition.\\ 


\begin{defini}($S_{Gen}^{EK}$: selection inside the protection of an external key)\label{PremierelemInt1}
$ $\\
We denote by $S_{Gen}^{EK}$ the class of all selections $S$ that meet the following conditions:\\
\\
$
\bullet ~S(\alpha,\alpha)={\mathcal{A}};$\begin{equation}\label{PremierelemInt1DefSgen1}\end{equation}
$\bullet ~S(\alpha,m)=\emptyset \mbox{,   if   } \alpha \not \in {\mathcal{A}}(m);$ \begin{equation}\label{PremierelemInt1DefSgen2}\end{equation}
$\bullet ~\forall \alpha \in {\mathcal{A}}(m),  \mbox { where } m=f_k(m_1, ..., m_n):$
\begin{equation}S(\alpha,m)  \subseteq (\underset{1\leq i \leq n}{\cup} {\mathcal{A}}(m_i){\cup} \{k^{-1}\}\backslash\{\alpha\})  \mbox{    if } f_k \in {\mathcal{E}}_{\mathcal{C}} \mbox{ and }\ulcorner {k^{-1}}\urcorner \sqsupseteq \ulcorner \alpha \urcorner \mbox{ and } m=m_\Downarrow\mbox{ } \label{PremierelemInt1DefSgen2222}\end{equation}
$\bullet ~\forall \alpha \in {\mathcal{A}}(m), \mbox { where } m=f(m_1, ..., m_n):$
\begin{equation}
S(\alpha,m)=    \begin{cases}
        \underset{1\leq i \leq n}{\cup} S(\alpha,m_i)& \mbox{if }  f_k \in {\mathcal{E}}_{\mathcal{C}} \mbox{ and } \ulcorner {k^{-1}}\urcorner \not \sqsupseteq \ulcorner \alpha \urcorner \mbox{ and } m=m_\Downarrow\mbox{ }  \mbox{   (a)}\\
        \underset{1\leq i \leq n}{\cup} S(\alpha,m_i) & \mbox{if } f\in \overline{{\mathcal{E}}}_{\mathcal{C}} \mbox{ and } m=m_\Downarrow \mbox{ }  \mbox{   (b)} \\
        S(\alpha,m_\Downarrow) &  \mbox{if } m\not=m_\Downarrow \mbox{   (c)}
    \end{cases}
\label{PremierelemInt1DefSgen3}
\end{equation}
$\bullet ~S(\alpha,\{m\}\cup M)=S(\alpha,m) \cup S(\alpha,M)$\begin{equation}
 \label{PremierelemInt1DefSgen4}
\end {equation}
\end{defini}

\noindent For an atom $\alpha$ in an encrypted message $m=f_k(m_1, ..., m_n)$, a selection $S$ as defined above returns a subset (see "$\subseteq$" in equation \ref{PremierelemInt1DefSgen2222}) among atoms that are neighbors of $\alpha$ in $m$ inside the protection of the most external protective key ${k}$ including its reverse form ${k^{-1}}$. The atom $\alpha$ itself is not selected. This set of candidate atoms is denoted by $\underset{1\leq i \leq n}{\cup} {\mathcal{A}}(m_i){\cup} \{k^{-1}\}\backslash\{\alpha\}$ in the equation \ref{PremierelemInt1DefSgen2222}. The most external protective key (or simply the external key) is the most external one that satisfies $\ulcorner {k^{-1}}\urcorner \sqsupseteq \ulcorner \alpha \urcorner$. A such key must exist when the set of messages generated by the protocol is well-protected, which is one of our assumptions above. 
By neighbor of $\alpha$ in $m$, we mean any atom that travels with it inside the protection of the external key. \\


\noindent $S_{Gen}^{EK}$ defines a generic class of selections since it does not identify what atoms to select precisely inside the protection of the external key. It identifies only the atoms that are candidates for selection and among them we are allowed to return any subset.\\ 

\begin{pre}\label{PremierelemBFSGENR}
$ $\\
Let $S\in$ $S_{Gen}^{EK}$ and ${\mathcal{C}}$ be a context of verification.\\
Let's have a rewriting system $\rightarrow_{\xi}$ such that $\forall m \in {\mathcal{M}},\forall \alpha \in {\mathcal{A}}(m) \wedge \alpha \not \in Clear(m),$ we have: \begin{equation} \forall l \rightarrow r \in \rightarrow_{\xi},S(\alpha,r) \subseteq S(\alpha,l)\label{eqRewSystSFIBIArt} \end{equation}
We have:
\center{$S$   is ${\mathcal{C}}$-reliable.}\\
\end{pre}

\begin{remarque}\label{RemRWSYSSFIBIart}
\noindent The condition on the rewriting system $\rightarrow_{\xi}$ given by the equation \ref{eqRewSystSFIBIArt} in the definition \ref{PremierelemBFSGENR} is introduced to make sure that the selection in the normal form is the smallest among all forms of a given message. This prevents the selection $S$ to select atoms that are inserted maliciously by the intruder by manipulating the equational theory. Hence, we are sure that all selected atoms by $S$ are honest. For example, let $m=\{\alpha.D\}_{k_{ab}}$ be a message in a homomorphic cryptography (i.e. $\{\alpha.D\}_{k_{ab}}=\{\alpha\}_{k_{ab}}.\{D\}_{k_{ab}}$). In the form $\{\alpha.D\}_{k_{ab}}$, the selection $S(\alpha,\{\alpha.D\}_{k_{ab}})$ may select $D$ since it is a neighbor of $\alpha$ inside the protection of $k_{ab}$, but in the form $\{\alpha\}_{k_{ab}}.\{D\}_{k_{ab}}$ the selection $S(\alpha,\{\alpha\}_{k_{ab}}.\{D\}_{k_{ab}})$ may not because it is not  a neighbor of $\alpha$. Then, we have to make sure that the rewriting system $\rightarrow_{\xi}$ we are using is oriented in such way that it chooses the form $\{\alpha\}_{k_{ab}}.\{D\}_{k_{ab}}$ rather than the form $\{\alpha.D\}_{k_{ab}}$ because there is no certitude that $D$ had not been inserted maliciously by the intruder using the homomorphic property in the equational theory. We assume that the rewriting system we are using meets this condition.\\
\end{remarque}

\noindent As for Proposition \ref{PremierelemBFSGENR}, 
it is easy to check that by construction a selection $S$, that is instance of $S_{Gen}^{EK}$, is well-formed. The proof of full-invariance-by-intruder is carried out by induction on the tree of construction of a message. The principal idea of the proof is that the selection related to an atom $\alpha$ in a message $m$ takes place inside the encryption by the most external protective key (such that: $\ulcorner k^{-1} \urcorner \sqsupseteq \ulcorner \alpha \urcorner$). Thus, an intruder cannot modify this selection when she does not have the key  $k^{-1}$ (i.e. $\ulcorner K(I) \urcorner \not \sqsupseteq  \ulcorner k^{-1}  \urcorner$). Besides, according to Lemma \ref{lematomique}, in a set of well-protected messages the intruder can never infer this key since it is atomic. So, this selection can only be modified  by people who are initially authorized to know $\alpha$ ( i.e.  $\ulcorner K \urcorner \sqsupseteq  \ulcorner k^{-1}  \urcorner$ and then $\ulcorner K \urcorner \sqsupseteq  \ulcorner \alpha \urcorner$). In addition, the intruder cannot neither use the equational theory to alter this selection thanks to the condition made on the rewriting system in the remark \ref{RemRWSYSSFIBIart}. Therefore any set of candidate atoms returned by $S$ cannot be altered (enlarged) by the intruder in any message $m$ that she can infer, as required by a full-invariant-by-intruder selection.\\
 

\begin{exemple}
$ $\\
Let $\alpha$ be an atomic message and $m$ a message such that: $\ulcorner \alpha \urcorner=\{A,B\}$ and $m=\{A.C.\alpha.D\}_{k_{ab}}$. 
Let $S_1, S_2$ and $S_3$ be three selections such that: $S_1(\alpha,m)=\{{k_{ab}^{-1}}\}$, $S_2(\alpha,m)=\{A, C,{k_{ab}^{-1}}\}$ and $S_3(\alpha,m)=\{A, C, D, {k_{ab}^{-1}}\}$. These three selections  are ${\mathcal{C}}$-reliable.
\end{exemple}

\subsection{Instantiation of reliable selections from the class $S_{Gen}^{EK}$}

\noindent Now that we defined a generic class of reliable selections $S_{Gen}^{EK}$, we will instantiate some concrete selections from it, that are naturally reliable. Instantiating $S_{Gen}^{EK}$ consists in defining selections that return precise sets of atoms among the candidates allowed by $S_{Gen}^{EK}$.
\subsubsection{The selection $S_{MAX}^{EK}$}
The selection $S_{MAX}^{EK}$ is the instance of the class $S_{Gen}^{EK}$ that returns for an atom in a message $m$ all its neighbors, that are principal identities, inside the protection of the external protective key $k$ in addition to its reverse key $k^{-1}$. (MAX means: the MAXimum of principal identities)
\vspace*{-0.5\baselineskip}
\subsubsection{The selection $S_{EK}^{EK}$}
The selection $S_{EK}^{EK}$ is the instance of the class $S_{Gen}^{EK}$ that returns for an atom in a message $m$ only the reverse key of the external protective key. (EK means: External Key)
\subsubsection{The selection $S_{N}^{EK}$}
The selection $S_{N}^{EK}$ is the instance of the class $S_{Gen}^{EK}$ that returns only its neighbors, that are principal identities, inside the protection of the external protective key. (N means: Neighbors)\\
\begin{exemple}
$ $\\
Let $\alpha$ be an atom and $m$ a message such that: $\ulcorner \alpha \urcorner = \{A, C\}$ and  $m=\{\{\{\alpha.E\}_{k_{ab}}.F\}_{k_{ac}}.D\}_{k_{ad}}$\\
$S_{MAX}^{EK}(\alpha,m)=\{E, F,k_{ac}^{-1}\};$ $S_{EK}^{EK}(\alpha,m)=\{k_{ac}^{-1}\};$ $S_{N}^{EK}(\alpha,m)=\{E, F\}$
\end{exemple}

\subsection{Specialized ${\mathcal{C}}$-reliable selection-based interpretation functions}
\noindent  Now, we define specific functions that are a composition of an appropriate homomorphism and instances of the class of selections $S_{Gen}^{EK}$. This homomorphism exports the properties of reliability from a selection to a function and transforms selected atoms to security levels. The following proposition states that any function that is a composition of the homomorphism defined below and the selections $S_{Gen}^{EK}$ is reliable.\\

\begin{pre}\label{preFCREK}
$ $\\
Let $\psi$ be a homomorphism defined as follows:
\begin{center}
\begin{tabular}{lrllll}
        &$\psi$&:&$({2^{\mathcal{A}}})^{\subseteq}$ &$\mapsto$& ${\mathcal{L}}^\sqsupseteq$ \\
        &&&$M$ &$\mapsto$&$\left\{
    \begin{array}{cl}   
        \top& \mbox{if } M=\emptyset  \\
         \underset{\alpha \in M}{\sqcap}\psi(\alpha) & \mbox{if not.}    
    \end{array}
\right.$\\
\\
&such that:&\multicolumn{4}{c}{$\psi(\alpha)=\left\{
    \begin{array}{ll}
        \{\alpha\} & \mbox{if } \alpha \in {\mathcal{I}} \mbox{ (Principal Identities) }  \\
        \ulcorner\alpha\urcorner & \mbox{if not.}
    \end{array}
\right.$}
\end{tabular}
\end{center}
\mbox{ We have: }$F_{MAX}^{EK}= \psi\circ S_{MAX}^{EK}$, $F_{EK}^{EK}=\psi\circ S_{EK}^{EK}$ and $F_{N}^{EK}=\psi\circ S_{N}^{EK}$ are  ${\mathcal{C}}$-reliable.\\
\end{pre}

\noindent The homomorphism $\psi$  in Proposition \ref{preFCREK} assigns for a principal in a selection, its identity. It assigns for a key its level of security in the context of verification. This homomorphism ensures the mapping from the operator "$\subseteq$"  to the operator "$\sqsupseteq$" in the lattice which offers to an interpretation function to inherit the full-invariance-by-intruder from its associated selection. In addition, it ensures the mapping from the operator "$\cup$"  to the operator "$\sqcap$" in the lattice, which offers to an interpretation function to be well-formed if its associated selection is well-formed.  Generally, every function $\psi\circ S$ remains reliable for any selection $S$ in $S_{Gen}^{EK}$.\\

\begin{exemple}
$ $\\
Let $\alpha$ be an atom, $m$ a message and $k_{ab}$ a key such that:\\
$\ulcorner \alpha \urcorner=\{A, B, S\}$; $m=\{A.C.\alpha.D\}_{k_{ab}}$; $\ulcorner{k_{ab}^{-1}}\urcorner=\{A, B, S\}$;\\
$S_{EK}^{EK}(\alpha,m)=\{{k_{ab}^{-1}}\}$; $S_{N}^{EK}(\alpha,m)=\{A,C,D\}$; $S_{MAX}^{EK}(\alpha,m)=\{A,C,D,{k_{ab}^{-1}}\}$;\\ 
$F_{EK}^{EK}(\alpha,m)=\psi\circ S_{EK}^{EK}(\alpha,m)=\ulcorner{k_{ab}^{-1}}\urcorner=\{A, B, S\}$;
$F_{N}^{EK}(\alpha,m)=\psi\circ S_{N}^{EK}(\alpha,m)=\{A, C, D\}$;
$F_{MAX}^{EK}(\alpha,m)=\psi\circ S_{MAX}^{EK}(\alpha,m)=\{A,C,D\}{ \cup}\ulcorner{k_{ab}^{-1}}\urcorner=\{A,C,D\} \cup \{A, B, S\}=\{A, C, D, B, S\}$.
\end{exemple}

%% file: ConfTemoinBorneeEnEK.tex
\section{\sc\textbf{Insufficiency of reliable function to analyze generalized roles}}\label{sectionWF}
\noindent So far, we presented a class of selection-based functions that have the required properties to analyze protocols statically. Unluckily, they operate  on valid traces that contain closed messages only. Nevertheless, a static  analysis must be led over the finite set of messages of the generalized roles of the protocol because the set of valid traces is infinite. The problem is that the finite set of  the generalized roles contains variables and the functions we defined are not "enough prepared" to analyze such messages because they are not supposed to be full-invariant by substitution~\cite{FranzRewriting,Rewriting,stabiliteParSubstitution}. The full-invariance by substitution is the property that allows us to perform an analysis over messages with variables and to export the conclusion made-on to closed messages. In the following section, we deal with the substitution question. We introduce the concept of derivative messages to reduce the impact of variables and we build the Witness-Functions that operate on these derivative messages rather than messages themselves. As we will see, the Witness-Functions provide two interesting  bounds that are independent of all substitutions. This fully replaces the property of full-invariance by substitution. Finally, we define a criterion of protocol correctness based on these two bounds.\\
\subsection{Derivative message}
\noindent Let $m,m_1,m_2$ $\in$ $\mathcal{M}$; $\mathcal{X}$$_m = Var(m)$; $S_1,S_2 \subseteq {2^{{\mathcal{X}}_m}}$; $\alpha \in$ ${\mathcal{A}}$$(m); X,Y \in {\mathcal{X}}_m$ and $\epsilon$ be the empty message.\\

\begin{defini}(Derivation)\label{derivation}
$ $\\
We define the derivative message as follows:
\begin{center}
\begin{tabular}{lll}
   $\partial_X \epsilon$ & $=$  $\epsilon$ &\\
   $\partial_X \alpha$ & $=$  $\alpha$ &\\
   $\partial_X X$ & $=$  $\epsilon$ &\\
   $\partial_X Y$ & $=$  $Y$, $X\neq Y$ &\\
   $\partial_X f(m)$ & $=$  $ f(\partial_X m), f\in {\mathcal{E}}_{\mathcal{C}} \cup \overline{{\mathcal{E}}}_{\mathcal{C}}$ &\\
   $\partial_{\{X\}} m$ & $=$  $\partial_{X} m$ &\\
   $\partial(\overline{X}) m$ & $=$  $\partial_{\{{\mathcal{X}}_m\backslash X\}} m$ &\\
   $\partial_{S_1 \cup S_2}m$ & $=$ $\partial_{S_2 \cup S_1}m$=  $\partial_{S_1}\partial_{S_2}m$=    
$\partial_{S_2}\partial_{S_1}m$&\\
\\
\end{tabular}
\end{center}
\end{defini}

\noindent To be simple, we denote by $\partial m$ the expression $\partial_{{\mathcal{X}}_m} m$. The operation of derivation introduced by the definition \ref{derivation} (denoted by $\partial$) eliminates variables in a message. $\partial_{X}m$ consists in eliminating the variable $X$ in $m$. $\partial(\overline{X})m$ consists in eliminating all variables, except $X$, in $m$. Hence, $X$ when overlined is considered as a constant in $m$. $\partial m$ consists in eliminating all the variables in $m$.\\

\begin{defini}\label{Fder}
$ $\\
Let $m\in {\mathcal{M}}_p^{\mathcal{G}}$, $X \in  {\mathcal{X}}_m$ and $m\sigma$ be a closed message.\\
For all $\alpha \in {\mathcal{A}}(m\sigma)$, $\sigma\in\Gamma$, we denote by:
\[
F(\alpha, \partial [\overline{\alpha}] m\sigma) = \left\{
    \begin{array}{ll}
        F(\alpha,\partial m) & \mbox{ if } \alpha \in {\mathcal{A}}(\partial m),\\
F(X,\partial [\overline{X}] m) & \mbox{ if } \alpha =X\sigma \wedge \alpha \notin {\mathcal{A}}(\partial m). \\

    \end{array}
\right.
\]
\end{defini}

\noindent A message $m$ in a generalized role is composed of two parts: a static part and a dynamic part. The dynamic part is described by variables. For an atom $\alpha$ in the static part (i.e. $\partial m$), $F(\alpha, \partial [\overline{\alpha}] m\sigma)$ removes the variables in $m$ and gives it the value $F(\alpha,\partial m)$. For anything that is not an atom of the static part -that comes by substitution of some variable $X$ in $m$- $F(\alpha, \partial [\overline{\alpha}] m\sigma)$ considers it as the variable itself, treated as a constant and as a block, and gives it all the time the same value: $F(X,\partial [\overline{X}] m)$. 
For any $F$ such that its associated selection is an instance of the class  $S_{Gen}^{EK}$, $F(\alpha, \partial [\overline{\alpha}] m\sigma)$ depends only on the static part of $m$ since $\alpha$ is not selected.
The function in the definition \ref{Fder} presents the three following major facts:\begin{enumerate}
  \item An atom of the static part of a message with variables, when analyzed with a such function, is considered as an atom in a message with no variables (a closed message);
  \item A variable, when analyzed by such function, is considered as any component that substitutes it (that is not in the static part of the message) with no respect to other variables, if any;
\item  For any $F$ such that its associated selection is an instance of the class  $S_{Gen}^{EK}$, $F(\alpha, \partial [\overline{\alpha}] m\sigma)$ depends only on the static part of $m$ since $\alpha$ is not selected.\\

\end{enumerate}

\noindent One could suggest that we attribute to an atom $\alpha$ in a closed message $m\sigma$ the value returned by the function $F(\alpha,\partial [\overline{\alpha}] m\sigma)$ given in the definition \ref{Fder} and hence we neutralize the variable effects. Unfortunately, this does not happen without undesirable "side-effects" because derivation generates a "loss of details". Let's look at the example \ref{exempleApp}.\\
\begin{exemple}\label{exempleApp}
$ $\\
Let $m_1$ and $m_2$ be two messages of a generalized role of a protocol $p$ such that 
$m_1=\{\alpha.C.X\}_{k_{ab}}$ and $m_2=\{\alpha.Y.D\}_{k_{ab}}$ and $\ulcorner\alpha\urcorner=\{A, B\}$;
\\ Let $m=\{\alpha.C.D\}_{k_{as}}$ be a closed message in a valid trace generated by $p$;\\
$$F_{MAX}^{EK}(\alpha,\partial [\overline{\alpha}] m)=\begin{cases}
              \{C, A, B\}& \mbox{if } m \mbox{ comes by the substitution of } X \mbox{ by }  D \mbox{ in }  m_1\\
              \{D, A, B\}& \mbox{if } m \mbox{ comes by the substitution of } Y \mbox{ by }  C \mbox{ in }  m_2\\
\end{cases}$$
Hence $F_{MAX}^{EK}(\alpha,\partial [\overline{\alpha}] m)$  is not even a function on the closed message $m$ since it may return more than one image for the same preimage. This leads us straightly to the Witness-Functions.
\end{exemple}
\section{The Witness-Functions}

\begin{defini}(Witness-Function)\label{WF}
$ $\\
Let $m\in {\mathcal{M}}_p^{\mathcal{G}}$, $X \in  {\mathcal{X}}_m$ and $m\sigma$ be a closed message.\\
Let $p$ be a protocol and $F$ be a ${\mathcal{C}}$-reliable function.\\
We define a Witness-Function ${\mathcal{W}}_{p,F}$ for all $\alpha \in {\mathcal{A}}(m\sigma)$, $\sigma\in\Gamma$, as follows: 
\[{\mathcal{W}}_{p,F}(\alpha,m\sigma)=\underset{\overset {m' \in {\mathcal{M}}_p^{\mathcal{G}}}{\exists \sigma' \in \Gamma.m'\sigma' = m \sigma }}{\sqcap} F(\alpha, \partial [\overline{\alpha}] m'\sigma')\]
${\mathcal{W}}_{p,F}$ is said to be a Witness-Function inside the protection of an external key when $F$ is afunction such that its associated selection is an instance of the class $S_{Gen}^{EK}.$\\
\end{defini}

\noindent According to the example \ref{exempleApp}, the application defined in \ref{Fder} is not necessary a function in  ${\mathcal{M}}_p^{\mathcal{G}}$ as a valid trace could have more than one source (or provenance) in ${\mathcal{M}}_p^{\mathcal{G}}$ and each source has a different static part. A Witness-Function is yet a function in ${\mathcal{M}}_p^{\mathcal{G}}$ since it searches all the sources of the closed message in input and returns the minimum (the union). This minimum naturally exists and is unique in the finite set ${\mathcal{M}}_p^{\mathcal{G}}$. A Witness-Function is protocol-dependent as it depends on messages in the generalized roles of the protocol. However, it is built uniformly for any pair (protocol, function) in input.  For a Witness-Function inside the protection of an external key, since its associated function calculates the level of security of an atom always in a message $m$ having an encryption pattern,
the search of the sources of the closed message $m\sigma$ in  ${\mathcal{M}}_p^{\mathcal{G}}$ (i.e. $\{{m' \in {\mathcal{M}}_p^{\mathcal{G}}| }{\exists \sigma' \in \Gamma.m'\sigma'=m\sigma}\}$) is reduced to a search in the encryption patterns in ${\mathcal{M}}_p^{\mathcal{G}}$.

\subsection{Legacy of Reliability}

\noindent Proposition \ref{WCReliableArtArticle} asserts that a function $F$ inside the protection of an external key transmits its reliability to its associated Witness-Function ${\mathcal{W}}_{p,F}$. In fact, the selection associated with a Witness-Function inside the protection of an external key is the union of selections associated with the function $F$, limited to derivative messages. It is easy to check that a Witness-Function is well-formed. Concerning the full-invariance-by-intruder property, as the derivation just eliminates  variables (so some atoms when the message is substituted), and since each selection in the union returns a subset among allowed candidates, then the union itself returns a subset among allowed candidates (the union of subsets is a subset). Therefore, the selection associated with a Witness-Function stays an instance of the class $S_{Gen}^{EK}$, so full-invariant-by-intruder. Since the Witness-Function is the composition of the homomorphism of $F$ and an instance of the class $S_{Gen}^{EK}$, then it is reliable. \\
\begin{pre}{}\label{WCReliableArtArticle}
$ $\\
Let ${\mathcal{W}}_{p,F}$ be a Witness-Function inside the protection of an external key.\\
We have:
\center{${\mathcal{W}}_{p,F}$ inherits reliability from $F$}
\end{pre}
\noindent\textit{See the proof 18 in~\cite{AppendixProofs}}\\

\begin{exemple}\label{exempleApp2}
$ $\\
Let  ${\mathcal{M}}_p^{\mathcal{G}}=\{\{\alpha.B.X\}_{k_{ad}},\{\alpha.Y.C\}_{k_{ad}}, \{A.Z\}_{k_{bc}}\};$   \\ $Var({\mathcal{M}}_p^{\mathcal{G}})=\{X, Y, Z\}$; 
$ m_1=\{\alpha.B.C\}_{k_{ad}}$; $\ulcorner\alpha\urcorner=\{A, D\}$;
$\ulcorner k_{ad}^{-1} \urcorner=\{A, D\}$;\\
\\
$~~~~{\mathcal{W}}_{p,F_{MAX}^{EK}}(\alpha,m_1)$\\
$={\mbox{ \{Definition \ref{WF}\}} }$\\
$ \hspace{1cm} \underset{\overset {m' \in {\mathcal{M}}_p^{\mathcal{G}}}{\exists \sigma' \in \Gamma.m'\sigma' = m_1}}{\sqcap} F_{MAX}^{EK}(\alpha, \partial [\overline{\alpha}] m'\sigma')=\underset{\overset {\{\{\alpha.B.X\}_{k_{ad}},\{\alpha.Y.C\}_{k_{ad}}\}}{\sigma'=\{X\longmapsto C,Y\longmapsto B\}}}{\sqcap} F_{MAX}^{EK}(\alpha, \partial [\overline{\alpha}] m'\sigma')$\\ 
$= \{ {\mathcal{W}}_{p,F_{MAX}^{EK}} {\mbox{ is well-formed from Proposition \ref{WCReliableArtArticle}\}} }$\\
$~~~~F_{MAX}^{EK}(\alpha, \partial [\overline{\alpha}] \{\alpha.B.X\}_{k_{ad}}[X\longmapsto C]) \sqcap$$F_{MAX}^{EK}(\alpha, \partial [\overline{\alpha}]\{\alpha.Y.C\}_{k_{ad}}[Y\longmapsto B])$\\
$=$${\mbox{\{Definition \ref{Fder} and derivation in \ref{derivation}\}}}$\\
$~~~~F_{MAX}^{EK}(\alpha, \{\alpha.B\}_{k_{ad}}) \sqcap F_{MAX}^{EK}(\alpha,\{\alpha.C\}_{k_{ad}})$\\
$={\mbox{\{Definition of }} F_{MAX}^{EK}\}$\\
$~~~~\{B,A,D\}\cup\{C,A,D\}=\{B,A,D,C\}$
\end{exemple}
\subsection{Bounds of a Witness-Function}

\noindent In Lemma \ref{corBoundsArtArticle}, we define two interresting bounds of a Witness-Function that are independent of all substitutions. The upper bound of a Witness-Function ranks the security level of an atom $\alpha$ in a closed message $m\sigma$ from one \textit{confirmed} source $m$ ($m$ is a natural source of $m\sigma$), the Witness-Function itself ranks it from the \textit{exact} sources of $m\sigma$  that are known only when the protocol is run, and the lower bound ranks it from all \textit{likely} sources of $m\sigma$ (i.e. the messages that are unifiable with $m$ in ${\mathcal{M}}_p^{\mathcal{G}}$). The unification in the lower bound catches all the principal identities that could participate in a possible intrusion. It is worth mentioning that the upper bound and the lower bound of the Witness-Function ${\mathcal{W}}_{p,F}$ do not depend on the substitution $\sigma$ that generates the closed message $m\sigma$.\\ 

\begin{lem}{}\label{corBoundsArtArticle}
$ $\\
Let $m\in{\mathcal{M}}_p^{\mathcal{G}}$ and ${\mathcal{W}}_{p,F}$ be a Witness-Function inside the protection of an external key.\\
$\forall \sigma \in \Gamma, \forall \alpha \in {\mathcal{A}}({\mathcal{M}}_p)$ we have:
\[F(\alpha,\partial [\overline{\alpha}] m) \sqsupseteq {\mathcal{W}}_{p,F}(\alpha,m\sigma) \sqsupseteq \underset{\overset {m' \in {\mathcal{M}}_p^{\mathcal{G}}}{\exists \sigma' \in \Gamma.m'\sigma' = m \sigma' }}{\sqcap} F(\alpha, \partial [\overline{\alpha}]m'\sigma')\]
\end{lem}


\begin{preuve2}{}\label{PRcorBoundsArticle}
$ $\\
For any $\sigma \in \Gamma$ we have:
\begin{itemize}
  \item [$\bullet$] $F(\alpha,\partial[\overline{\alpha}] m)\subseteq {\mathcal{W}}_{p,F}(\alpha,m\sigma)$: 
since $m$ is obviously one element of the set $\{m' \in {\mathcal{M}}_p^{\mathcal{G}}|\exists \sigma' \in \Gamma.m'\sigma' = m \sigma\}$ of calculation of ${\mathcal{W}}_{p,F}(\alpha,m\sigma)$  (i.e. $m$ is a trivial source of $m\sigma$) 
and since $ F(\alpha,\partial[\overline{\alpha}] m\sigma)$ does not depend on $\sigma$ because, by construction, it depends only on the static part of $m$ (denoted simply by $F(\alpha,\partial[\overline{\alpha}] m)$).
  \item [$\bullet$] ${\mathcal{W}}_{p,F}(\alpha,m\sigma)\subseteq \underset{\overset {m' \in {\mathcal{M}}_p^{\mathcal{G}}}{\exists \sigma' \in \Gamma.m'\sigma' = m \sigma' }}{\cup} F(\alpha, \partial [\overline{\alpha}]m'\sigma')$: 
since for all $m \in {\mathcal{M}}_p^{\mathcal{G}}$ the set $\{m' \in {\mathcal{M}}_p^{\mathcal{G}}|\exists \sigma' \in \Gamma.m'\sigma' = m \sigma'\}$ (unifications) is obviously larger than the set $\{m' \in {\mathcal{M}}_p^{\mathcal{G}}|\exists \sigma' \in \Gamma.m'\sigma' = m \sigma\}$ of sources of $m\sigma$ in ${\mathcal{M}}_p^{\mathcal{G}}$.\\
From these two facts and since ${\mathcal{L}}^\sqsupseteq$ is a lattice, we have the result in Lemma \ref{corBoundsArtArticle}
\end{itemize}
\end{preuve2}


\section{\sc\textbf{A Semi-Decidable Procedure for Analyzing Cryptographic Protocols with the Witness-Functions}}


\noindent Now, we give the protocol analysis with a Witness-Function theorem that sets a criterion for protocols correctness with respect to the secrecy property. The result in Theorem \ref{PAT} derives directly from Proposition \ref{WCReliableArtArticle}, Lemma \ref{corBoundsArtArticle} and Theorem \ref{mainThArticle}.\\ 

\begin{thm}(Protocol analysis with a Witness-Function)\label{PAT}
$ $\\
Let ${\mathcal{W}}_{p,F}$ be a Witness-Function inside the protection of an external key.\\
A sufficient condition of correctness of $p$ with respect to the secrecy property is:\\
$\forall R.r \in R_G(p), \forall \alpha \in {\mathcal{A}}{(r^+ )}$ we have:\\
\[\underset{\overset {m' \in {\mathcal{M}}_p^{\mathcal{G}}}{\exists \sigma' \in \Gamma.m'\sigma' = r^+ \sigma' }}{\sqcap} F(\alpha, \partial[\overline{\alpha}] m'\sigma') \sqsupseteq \ulcorner \alpha \urcorner \sqcap F(\alpha,\partial[\overline{\alpha}] R^-)\]\end{thm}



\begin{preuve2}{}\label{PATArticle}
$ $\\
$\mbox{ Suppose we have:                         }\forall R.r \in R_G(p), \forall \alpha \in {\mathcal{A}}{(r^+ )}$
\begin{equation}
\underset{\overset {m' \in {\mathcal{M}}_p^{\mathcal{G}}}{\exists \sigma' \in \Gamma.m'\sigma' = r^+ \sigma' }}{\sqcap} F(\alpha, \partial[\overline{\alpha}] m'\sigma') \sqsupseteq \ulcorner \alpha \urcorner \sqcap F(\alpha,\partial [\overline{\alpha}]R^-) \label{LapreuvDuDernier4Article}
\end{equation}
From Lemma \ref{corBoundsArtArticle} and since  ${\mathcal{L}}^\sqsupseteq$ is a lattice we have for all $\sigma \in \Gamma$:
\begin{equation}
\forall \alpha \in {\mathcal{A}}({\mathcal{M}}_p).{\mathcal{W}}_{p,F}(\alpha, r^+\sigma )\sqsupseteq\underset{\overset {m' \in {\mathcal{M}}_p^{\mathcal{G}}}{\exists \sigma' \in \Gamma.m'\sigma' = r^+ \sigma' }}{\sqcap} F(\alpha, \partial [\overline{\alpha}]m'\sigma') \label{LapreuvDuDernier2Article}\end{equation}
\begin{equation}
\mbox{and }\forall \alpha \in {\mathcal{A}}({\mathcal{M}}_p).\ulcorner \alpha \urcorner \sqcap F(\alpha,\partial[\overline{\alpha}] R^-) \sqsupseteq  \ulcorner \alpha \urcorner \sqcap {\mathcal{W}}_{p,F}(\alpha,R^-\sigma )   \label{LapreuvDuDernier3Article}\end{equation}
From \ref{LapreuvDuDernier4Article}, \ref{LapreuvDuDernier2Article} and \ref{LapreuvDuDernier3Article} we have:
\begin{equation}
\forall \alpha \in {\mathcal{A}}(r^+\sigma).{\mathcal{W}}_{p,F}(\alpha, r^+\sigma )\sqsupseteq \ulcorner \alpha \urcorner \sqcap {\mathcal{W}}_{p,F}(\alpha,R^-\sigma )\label{LapreuvDuDernier1Article}\end{equation}
From Proposition \ref{WCReliableArtArticle}  ${\mathcal{W}}_{p,F}$ is  ${\mathcal{C}}$-reliable, then we have from Theorem \ref{PR5Article} and \ref{LapreuvDuDernier1Article}: 
\[p \mbox{ is correct with respect to the secrecy property}\Box\]
\end{preuve2}


The secrecy in cryptographic protocols is in general an undecidable problem~\cite{Combinaison-Eq-journal,CortierSurvey1,Durgin1,Lundh1,Ref42}. Theorem \ref{PAT} defines a semi-decidable procedure for analyzing cryptographic protocols with the Witness-Functions for the class of increasing protocols.  Hence, the problem of  protocol correctness becomes a problem of protocol growth. If the criterion in Theorem \ref{PAT} is met, we can conclude that the protocol is correct for secrecy. Otherwise, no decision can be made. This criterion uses the upper bound and the lower bound of a Witness-Function. The upper bound gives a reasonable estimation of the level of security of an atom in a message on its reception. This estimation is based on selection among the external protective key and the principal identities that travel with it under the same protection. The lower bound makes sure that no unauthorized principal identity is maliciously inserted beside this atom on sending and therefore ensures that no intrusion could happen. The fact that the two bounds of the Witness-Function are not affected by substitution, our semi-decidable procedure can transfer any decision made on the generalized roles of a protocol to the valid traces and with no restriction on their size.\\

\begin{center}
\begin{figure}[ht!]
\begin{center}
\includegraphics[scale=0.50]{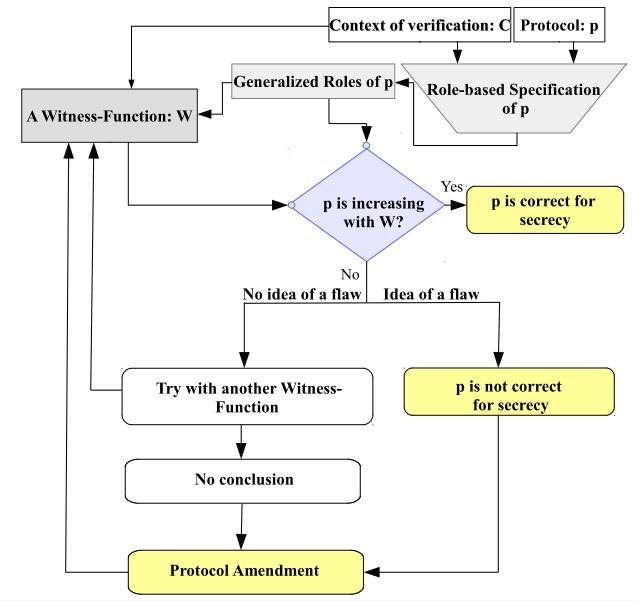}
\end{center}
\caption{Analyzing cryptographic protocols with Witness-Functions}
\label{AnCryPrWF}
\end{figure}
\end{center}

If an analysis with a Witness-Function fails to prove the correctness of a protocol for secrecy, we can change the Witness-Function by changing its function in entry. We can also have an idea about some flaw in it. If all the Witness-Functions that we manage to build fail, we can advice an amendment to the structure of the protocol by changing some encryption keys or principal identities in the exchanged messages in the protocol. 
The procedure steps are given in Figure \ref{AnCryPrWF}.

\vspace*{-0.5\baselineskip}

%% file: AnalyseNS.tex
\section{\sc\textbf{Analysis of the Needham-Schroeder protocol with a Witness- Function}}\label{sectionAn2}

\noindent In this section, we analyze the Needham-Schroeder protocol  given in Example \ref{exNS} with a Witness-Function.

\noindent Let us have a context of verification such that: $\ulcorner A\urcorner= \bot$; $\ulcorner B\urcorner= \bot$; $\ulcorner N_a^i\urcorner= \{A,B\}$ (secret between $A$ and $B$); $\ulcorner N_b^i\urcorner= \{A,B\}$ (secret between $A$ and $B$); $\ulcorner k_a^{-1}\urcorner=\{A\}$; $\ulcorner k_b^{-1}\urcorner=\{B\}$; $({\mathcal{L}},\sqsupseteq, \sqcup, \sqcap, \bot,\top)=(2^{\mathcal{I}},\subseteq,\cap,\cup,\mathcal{I}, \emptyset)$;  
${\mathcal{I}}=\{I(intruder), A, B, C, A_1, A_2, B_1, B_2,...\}$;\\
\\
\noindent The set of messages generated by the protocol is ${\mathcal{M}}_p^{\mathcal{G}}=$ 
$\{\{A_1.N_{A_1}\}_{k_{B_1}},\{N_{A_2}.X_1.B_2\}_{k_{A_2}},\\ \{X_2\}_{k_{B_3}}, 
\{A_3.Y_1\}_{k_{B_4}},\{Y_2.N_{B_5}.B_5\}_{k_{A_4}}, $ $  \{N_{B_6}\}_{k_{B_6}} \}$;\\
\noindent The variables are denoted by $X_1, X_2, Y_1$ and $Y_2$;\\ 
\noindent The static names are denoted by $A_1, N_{A_1}, k_{B_1},
N_{A_2}, B_2, k_{A_2}, k_{B_3}, A_3, k_{B_4}, N_{B_5}, B_5, k_{A_4}, N_{B_6}$ and $k_{B_6}$;\\
\\
\noindent Let's select the Witness-Function as follows:\\ 
$p=NS$; $F=F_{MAX}^{EK}$; ${\mathcal{W}}_{p,F}(\alpha,m\sigma)=\underset{\overset {m' \in {\mathcal{M}}_p^{\mathcal{G}}}{\exists \sigma' \in \Gamma.m'\sigma' = m\sigma }}{\sqcap} F(\alpha, \partial[\overline{\alpha}] m'\sigma')$;\\
Let's denote the lower bound of the Witness-Function by: $${\mathcal{W}}'_{p,F}(\alpha,r^+)=\underset{\overset {m' \in {\mathcal{M}}_p^{\mathcal{G}}}{\exists \sigma' \in \Gamma.m'\sigma' = r^+\sigma' }}{\sqcap} F(\alpha, \partial [\overline{\alpha}]m'\sigma')$$

\subsubsection{Analysis of the generalized role of $A$}
$ $\\
\\
As defined in the generalized roles of $p$, an agent $A$ can participate in two consequent sessions: $S^{i}$ and $S^{j}$ such that $j>i$. In the former session $S^{i}$, the agent $A$ receives nothing and sends the message $\{A.N_a^i\}_{k_b}$. In a consequent session $S^{j}$, she receives the message $\{N_a^i.X.B\}_{k_a}$ and she sends the message $\{X\}_{k_b}$. This is described by the following schema:
$${S^{i}}:\frac{\Box}{\{A.N_a^i\}_{k_b} }~~~~~~~~~~~~~~~~~{S^{j}}:\frac{\{N_a^i.X.B\}_{k_a}}{\{X\}_{k_b}}$$
\textbf{-Analysis of the messages exchanged  in the session $S^{i}$:}\\
\\
1- For any $N_a^i$:\\
a- When sending: $r_{S^i}^+=\{A.N_a^i\}_{k_b}$\textit{(in a sending step, the lower bound is used)}\\
 $\forall N_a^i.\{m' \in {\mathcal{M}}_p^{\mathcal{G}}|{\exists \sigma' \in \Gamma.m'\sigma' = r_{S^i}^+\sigma' } \}$\\$=\forall N_a^i.\{m' \in {\mathcal{M}}_p^{\mathcal{G}}|{\exists \sigma' \in \Gamma.m'\sigma' = \{A.N_a^i\}_{k_b}\sigma' } \}$ \\$=\{(\{A_1.N_{A_1}\}_{k_{B_1}},\sigma_1'), (\{X_2\}_{k_{B_3}},\sigma_2'), (\{A_3.Y_1\}_{k_{B_4}},\sigma_3')\}$ such that: 
$$\left\{
    \begin{array}{l}
      \sigma_1'=\{A_1 \longmapsto A, N_{A_1} \longmapsto N_a^i, k_{B_1} \longmapsto k_b\}  \\
      \sigma_2'=\{ X_2 \longmapsto A.N_a^i, k_{B_3} \longmapsto k_b\} \\
      \sigma_3'=\{ A_3 \longmapsto A, Y_1 \longmapsto N_a^i, k_{B_4} \longmapsto {k_b} \}
    \end{array}
\right.$$
${\mathcal{W}}'_{p,F}(N_a^i,\{A.N_a^i\}_{k_b})$\\
$=\{\mbox{Definition of the lower bound of the Witness-Function}\}$\\
$F(N_a^i,\partial [\overline{N_a^i}]\{A_1.N_{A_1}\}_{k_{B_1}} \sigma_{1}') \sqcap F(N_a^i,\partial [\overline{N_a^i}]\{X_2\}_{k_{B_3}} \sigma_{2}') \sqcap$$ F(N_a^i,\partial[\overline{N_a^i}] \{A_3.Y_1\}_{k_{B_4}} \sigma_{3}')$\\
$=\{\mbox{Setting the static neighborhood}\}$\\
$F(N_a^i,\partial [\overline{N_a^i}]\{A.N_a^i\}_{k_{b}} \sigma_{1}') \sqcap F(N_a^i,\partial [\overline{N_a^i}]\{X_2\}_{k_{b}} \sigma_{2}') \sqcap$$ F(N_a^i,\partial[\overline{N_a^i}] \{A.Y_1\}_{k_{b}} \sigma_{3}')$\\
$=\{\mbox{Definition } \ref{Fder}\}$
\\
$ F(N_a^i,\{A.N_a^i\}_{k_{b}}) \sqcap F(X_2,\partial [\overline{X_2}]\{X_2\}_{k_{b}}) \sqcap$$ F(Y_1,\partial[\overline{Y_1}] \{A.Y_1\}_{k_{b}})$\\
$=\{\mbox{Derivation in Definition } \ref{derivation}\}$
\\
$F(N_a^i, \{A.N_a^i\}_{k_b}) \sqcap F(X_2, \{X_2\}_{k_b}) \sqcap$$ F(Y_1, \{A.Y_1\}_{k_b})$\\
$=\{\mbox{Since } F=F_{MAX}^{EK}\}$\\
$\{A,B\} \cup \{B\}  \cup \{A,B\}=\{A,B\}$ (1.0)\\
\\
b- When receiving: $R_{S^i}^-=\emptyset$ \textit{(in a receiving step, the upper bound is used)}\\
$F(N_a^i,\emptyset)=\top$ (1.1)\\
\\
2- Conformity to the criterion set by Theorem \ref{PAT} of the exchanged messages in the session $S^{i}$:\\
From (1.0) and (1.1), we have:\\
${\mathcal{W}}'_{p,F}(N_a^i,\{A.N_a^i\}_{k_b})= \{A,B\} \sqsupseteq \ulcorner N_a^i \urcorner \sqcap F(N_a^i,\emptyset)=\{A,B\}\sqcap\top= \{A,B\}$ (1.2)\\
From (1.2), the messages exchanged in the session $S^{i}$ respect the  criterion set by Theorem \ref{PAT}. (I)\\
\\
\textbf{-Analysis of the messages exchanged  in the session $S^{j}$:}\\
\\
1- For any  $X$:\\
a- When sending: $R_{S^j}^+=\{X\}_{k_b}$ \textit{(in a sending step, the lower bound is used)}\\
$\forall X.\{m' \in {\mathcal{M}}_p^{\mathcal{G}}|{\exists \sigma' \in \Gamma.m'\sigma'= R_{S^j}^+\sigma' } \}$
$=\forall X.\{m' \in {\mathcal{M}}_p^{\mathcal{G}}|{\exists \sigma' \in \Gamma.m'\sigma' = \{X\}_{k_b}\sigma' } \}$\\
$=\{ (\{X_2\}_{k_{B_3}},\sigma_1') \}$ such that: $\sigma_1'=\{X_2 \longmapsto X, k_{B_3} \longmapsto k_b \}$.\\
\\
${\mathcal{W}}'_{p,F}(X,\{X\}_{k_b})$\\
$=\{\mbox{Definition of the lower bound of the Witness-Function}\}$\\
$F(X,\partial [\overline{X}]\{X_2\}_{k_{B_3}}\sigma_{1}')$\\
$=\{\mbox{Setting the static neighborhood}\}$\\
$F(X,\partial [\overline{X}]\{X_2\}_{k_{b}}\sigma_{1}')$\\
$=\{\mbox{Definition } \ref{Fder}\}$\\
$F(X_2, \partial [\overline{X_2}] \{X_2\}_{k_{b}})$\\
$=\{\mbox{Derivation in Definition } \ref{derivation}\}$\\
$F(X_2, \{X_2\}_{k_{b}})$\\
$=\{\mbox{Since } F=F_{MAX}^{EK}\}$\\
$\{B\}$ (2.0)\\
\\
b- When receiving: $R_{S^j}^-=\{N_a^i.X.B\}_{k_a}$   \textit{(in a receiving step, the upper bound is used)}\\
$F(X,\partial[\overline{X}]\{N_a^i.X.B\}_{k_a})=$
$\{A,B\}$ (2.1)\\
\\
3- Conformity to the  criterion set by Theorem \ref{PAT} of the exchanged messages in the session $S^{j}$:\\
From (2.0) et (2.1), we have:\\
${\mathcal{W}}'_{p,F}(X,\{X\}_{k_b})=\{B\}  \sqsupseteq \ulcorner X \urcorner \sqcap F(X,\partial [\overline{X}] \{N_a^i.X.B\}_{k_a})=\ulcorner X \urcorner \cup \{A,B\}$ (2.2)\\
From (2.2), the messages exchanged in the session $S^{j}$ respect the criterion set by Theorem \ref{PAT}. (II)\\
From (I) and (II), the messages exchanged in the generalized role of $A$ respect the  criterion set by Theorem \ref{PAT}. (III)

\subsubsection{Analysis of the generalized role of $B$}
$ $\\
\\
As defined in the generalized roles of $p$, an agent $B$ can participate in a session ${{S'}^{^{i}}}$, in which she receives the message $\{A.Y\}_{k_b}$ and she sends the message $\{Y.N_b^i.B\}_{k_a}$. This is described by the following schema:
\[{{{S'}^{^{i}}}}:\frac{\{A.Y\}_{k_b}}{\{Y.N_b^i.B\}_{k_a}}
\]
\\
\textbf{-Analysis of the messages exchanged  in the session ${{S'}^{^{i}}}$:}\\
\\
1- For any $Y$:\\
a- When sending: $r_{{{S'}^{^{i}}}}^+=\{Y.N_b^i.B\}_{k_a}$\textit{(in a sending step, the lower bound is used)}\\
$\forall Y.\{m' \in {\mathcal{M}}_p^{\mathcal{G}}|{\exists \sigma' \in \Gamma.m'\sigma' = r_{{{S'}^{^{i}}}}^+\sigma' }\}$\\
$=\forall Y.\{m' \in {\mathcal{M}}_p^{\mathcal{G}}|{\exists \sigma' \in \Gamma.m'\sigma' =\{Y.N_b^i.B\}_{k_a}\sigma' }\}$\\
$=\{ (\{Y_2.N_{B_5}.B_5\}_{k_{A_4}},\sigma_1'), (\{X_2\}_{k_{B_3}},\sigma_2')\}$ such that: $$\left\{
    \begin{array}{l}
      \sigma_1'=\{Y_2 \longmapsto Y, N_{B_5} \longmapsto N_b^i, B_5 \longmapsto B, k_{A_4} \longmapsto k_a\}\\
      \sigma_2'=\{X_2 \longmapsto Y.N_b^i.B, k_{B_3} \longmapsto k_a\}
    \end{array}
\right.$$
${\mathcal{W}}'_{p,F}(Y,\{Y.N_b^i.B\}_{k_a})$\\
$=\{\mbox{Definition of the lower bound of the Witness-Function}\}$\\
$F(Y,\partial[\overline{Y}] \{Y_2.N_{B_5}.B_5\}_{k_{A_4}} \sigma_{1}') \sqcap F(Y,\partial[\overline{Y}] \{X_2\}_{k_{B_3}} \sigma_{2}')$\\
$=\{\mbox{Setting the static neighborhood}\}$\\
$F(Y,\partial[\overline{Y}] \{Y_2.N_b^i.B\}_{k_{a}} \sigma_{1}') \sqcap F(Y,\partial[\overline{Y}] \{X_2\}_{k_{a}} \sigma_{2}')$\\
$=\{\mbox{Definition } \ref{Fder}\}$\\
$F(Y_2,\partial[\overline{Y_2}] \{Y_2.N_b^i.B\}_{k_{a}}) \sqcap F(X_2,\partial[\overline{X_2}] \{X_2\}_{k_{a}})$\\
$=\{\mbox{Derivation in Definition } \ref{derivation}\}$\\
$F(Y_2,\{Y_2.N_b^i.B\}_{k_{a}}) \sqcap F(X_2,\{X_2\}_{k_{a}})$\\
$=\{\mbox{Since } F=F_{MAX}^{EK}\}$\\
$\{A,B\}\cup \{A\}=\{A, B\}$ (3.0)\\
\\
b- When receiving: $R_{{{S'}^{^{i}}}}^-=\{A.Y\}_{k_b}$ \textit{(in a receiving step, the upper bound is used)}\\
$F(Y,\partial[\overline{Y}]\{A.Y\}_{k_b})=$
$\{A,B\}$(3.1)\\
\\
2- For any $N_b^i$:\\
a- When sending: $r_{{{S'}^{^{i}}}}^+=\{Y.N_b^i.B\}_{k_a}$\textit{(in a sending step, the lower bound is used)}\\
$\forall N_b^i.\{m' \in {\mathcal{M}}_p^{\mathcal{G}}|{\exists \sigma' \in \Gamma.m'\sigma' = r_{{{S'}^{^{i}}}}^+\sigma' } \}$\\
$=\forall N_b^i.\{m' \in {\mathcal{M}}_p^{\mathcal{G}}|{\exists \sigma' \in \Gamma.m'\sigma' = \{Y.N_b^i.B\}_{k_a}\sigma' } \}$\\
$=\{ (\{Y_2.N_{B_5}.B_5\}_{k_{A_4}},\sigma_1'), (\{X_2\}_{k_{B_3}},\sigma_2'), (\{A_3.Y_1\}_{k_{B_4}},\sigma_3') \}$
 such that: $$\left\{
    \begin{array}{l}
      \sigma_1'=\{Y_2 \longmapsto Y, N_{B_5} \longmapsto N_b^i, B_5 \longmapsto B, k_{A_4} \longmapsto k_a\}\\
      \sigma_2'=\{X_2 \longmapsto Y.N_b^i.B, k_{B_3} \longmapsto k_a\}\\
      \sigma_3'=\{Y\longmapsto A_3, Y_1 \longmapsto N_b^i.B, k_{B_4} \longmapsto {k_a}\}
    \end{array}
\right.$$
${\mathcal{W}}'_{p,F}(N_b^i,\{Y.N_b^i.B\}_{k_a})$\\
$=\{\mbox{Definition of the lower bound of the Witness-Function}\}$\\
$F(N_b^i, \partial [\overline{N_b^i}] \{Y_2.N_{B_5}.B_5\}_{k_{A_4}}\sigma_{1}'  ) \sqcap$$ F(N_b^i, \partial [\overline{N_b^i}] \{X_2\}_{k_{B_3}}\sigma_{2}') \sqcap F(N_b^i, \partial [\overline{N_b^i}]\{A_3.Y_1\}_{k_{B_4}} \sigma_{3}')$\\
$=\{\mbox{Setting the static neighborhood}\}$\\
$F(N_b^i, \partial [\overline{N_b^i}] \{Y_2.N_b^i.B\}_{k_{a}}\sigma_{1}'  ) \sqcap$\\$ F(N_b^i, \partial [\overline{N_b^i}] \{X_2\}_{k_{a}}\sigma_{2}') \sqcap F(N_b^i, \partial [\overline{N_b^i}]\{A_3.Y_1\}_{k_{a}} \sigma_{3}')$\\
$=\{\mbox{Definition } \ref{Fder}\}$\\
$F(N_b^i, \{N_b^i.B\}_{k_{a}}) \sqcap$$ F(X_2, \partial [\overline{X_2}] \{X_2\}_{k_{a}}) \sqcap F(Y_1, \partial [\overline{Y_1}]\{A_3.Y_1\}_{k_{a}})$\\
$=\{\mbox{Derivation in Definition } \ref{derivation}\}$\\
$F(N_b^i, \{N_b^i.B\}_{k_{a}}) \sqcap$$ F(X_2,  \{X_2\}_{k_{a}}) \sqcap F(Y_1,\{A_3.Y_1\}_{k_{a}})$\\
$=\{\mbox{Since } F=F_{MAX}^{EK}\}$\\
$\{A,B\} \cup \{A\} \cup \{A_3,A\}=\{A,B,A_3\}$ (3.2)\\
\\
b- When receiving: $R_{{{S'}^{^{i}}}}^-=\{A.Y\}_{k_b}$ \textit{(in a receiving step, the upper bound is used)}\\
$F(N_b^i,\partial[\overline{N_b^i}] \{A.Y\}_{k_b})=\top$ (3.3)\\
\\
3- Conformity to the  criterion set by Theorem \ref{PAT} of the exchanged messages in the session ${{S'}^{^{i}}}$:\\
From (3.0) et (3.1), we have:\\
${\mathcal{W}}'_{p,F}(Y,\{Y.N_b^i.B\}_{k_a})= \{A,B\} \sqsupseteq \ulcorner Y \urcorner \sqcap  F(Y,\partial [\overline{Y}] \{A.Y\}_{k_b})=\ulcorner Y \urcorner \cup \{A,B\}$(3.4)\\
From (3.2) et (3.3), we have:\\
${\mathcal{W}}'_{p,F}(N_b^i, \{Y.N_b^i.B\}_{k_a})=\{A,B,A_3\} \not \sqsupseteq \ulcorner N_b^i \urcorner \sqcap F(N_b^i,\partial [\overline{N_b^i}]\{A.Y\}_{k_b})=\ulcorner N_b^i \urcorner \sqcap \top=\ulcorner N_b^i \urcorner =\{A,B\}$ (3.5). From (3.5), the messages exchanged in the session ${{S'}^{^{i}}}$ do not respect the  criterion set by Theorem \ref{PAT}. (IV)

\subsection{Results and Interpretation}
\noindent The results of the analysis of the variation of Needham-Schroeder protocol are summarized in Table \ref{NSGrowthNSLlong}.

\begin{center}
\begin{table}[h]
\caption[]{\label{NSGrowthNSLlong} Conformity of the Needham-Schroeder Protocol to Theorem \ref{PAT}}
\begin{center}
\begin{tabular}{|c|c|c|c|c|c|c|c|}
  \hline
  $\alpha$ &  Role & $R^-$ & $r^+$ & ${\mathcal{W}}'_{p,F}(\alpha, r^+)$& $\ulcorner \alpha \urcorner$ & $ F(\alpha,\partial[\overline{\alpha}] R^-)$& Theorem \ref{PAT}\\
  \hline
  $N_a^i$ & $A$& $\emptyset$ & $\{A.N_a^i\}_{k_b}$ & $\{A, B\}$& $\{A, B\}$ &  $ \top$& Fulfilled\\
  \hline
 $\forall X$ & $A$ & $\{N_a^i.X.B\}_{k_a}$ & $\{X\}_{k_b}$ & $\{B\}$& $\ulcorner X \urcorner$ &  $ \{A, B\}$& Fulfilled\\
  \hline
 \hline
 $\forall Y$ & $B$ & $\{A.Y\}_{k_b}$ & $\{Y.N_b^i.B\}_{k_a}$ & $\{A, B\}$& $\ulcorner Y \urcorner$ &  $ \{A, B\}$& Fulfilled\\
 \hline
  $N_b^i$ & $B$ & $\{A.Y\}_{k_b}$ & $\{Y.N_b^i.B\}_{k_a}$ & $\{A, B, A_3\}$&  $\{A, B\}$ & $ \{A, B\}$& Not Fulfilled\\
   \hline 
\end{tabular}
\end{center}
\end{table}
\end{center}

\noindent We notice from Table \ref{NSGrowthNSLlong} that the variation of Needham-Schroeder protocol does not respect the  criterion set by Theorem \ref{PAT} when analyzed with the witness-function  ${\mathcal{W}}_{{p_{{_{NS}}}},F_{MAX}^{EK}}$. Therefore, we cannot deduce anything regarding its correctness with respect to the secrecy property. The non-growth of the protocol is localized in the sending step of the generalized role of $B$ and it is due to a  possible malicious neighbor (denoted by $A_3$ in our analysis) that could be inserted beside the nonce $N_{B}^i$. In the literature, we report a flaw that operates on the decay of the level of security of the nonce $N_{B}^i$ in the generalized role of $B$. This flaw is described by the attack scenario in Figure.\ref{FailleNSDescAnnexeLong}.
\vspace*{-1.0\baselineskip}

\small{
\setlength{\unitlength}{1mm}
\thicklines
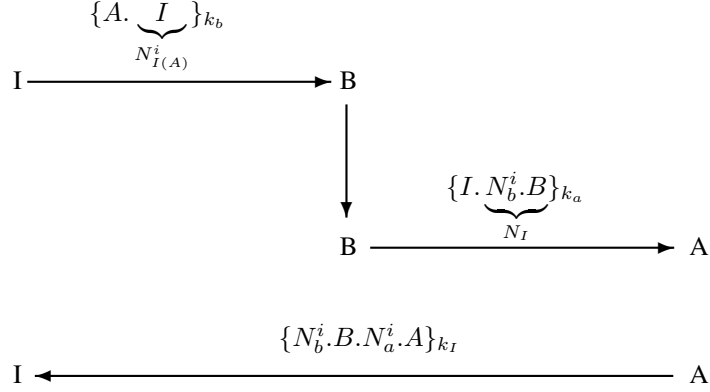
\begin{figure}[h]
   \centering

\begin{picture}(40,20)(25,0)
\put(0,-1){I}
\put(10,8){$\{A.\underbrace{I}_{N_{I(A)}^i}\}_{k_b}$}
\put(1,0){
   \vector(1,0){40}
}
\put(43,-1){B}
\put(43,-3){
   \vector(0,-1){15}
}
\put(43,-23){B}
\put(47,-22){\vector(1,0){40}}
\put(57,-15){$\{I.\underbrace{N_b^i.B}_{N_{I}}\}_{k_a}$}
\put(89,-23){A}
\put(89,-40){A}
\put(86,-39){
   \vector(-1,0){84}
}
\put(35,-35){$\{N_b^i.B.N_a^i.A\}_{k_{I}}$}

\put(0,-40){I}
\end{picture}
\\\hspace{\linewidth}
\\\hspace{\linewidth}
\\\hspace{\linewidth}
\\\hspace{\linewidth}
\\\hspace{\linewidth}
\\\hspace{\linewidth}
\\\hspace{\linewidth}
\\\hspace{\linewidth}
\\\hspace{\linewidth}
\\\hspace{\linewidth}
\caption[]{\label{FailleNSDescAnnexeLong} Attack Scenario on the Needham-Schroeder Protocol}
\end{figure}
}
\normalsize

%% file: Exemple2Leger.tex
 \vspace*{-1.5\baselineskip}
\section{\sc\textbf{Analysis of the NSL protocol with a Witness-Function}}\label{sectionAn1}

\noindent In this section, we analyze the NSL protocol with a witness-function given in Example \ref{exNSL}.
\noindent Let's have a context of verification such that: $\ulcorner A\urcorner= \bot$; $\ulcorner B\urcorner= \bot$; $\ulcorner N_a^i\urcorner= \{A,B\}$; $\ulcorner N_b^i\urcorner= \{A,B\}$; $\ulcorner k_a^{-1}\urcorner=\{A\}$; $\ulcorner k_b^{-1}\urcorner=\{B\}$; $({\mathcal{L}},\sqsupseteq, \sqcup, \sqcap, \bot,\top)=(2^{\mathcal{I}},\subseteq,\cap,\cup,\mathcal{I}, \emptyset)$;
${\mathcal{I}}=\{I, A, B, A_1, A_2, $$B_1, B_2,...\}$;
\\
\noindent The set of messages generated by the protocol is ${\mathcal{M}}_p^{\mathcal{G}}=$
$\{\{N_{A_1}.A_1\}_{k_{B_1}},
\{B_2.N_{A_2}\}_{k_{A_2}},
\{B_3.X_1\}_{k_{A_3}},\\
\{X_2\}_{k_{B_4}},
\{Y_1.A_{4}\}_{k_{B_5}},
\{B_7.N_{B_7}\}_{k_{A_6}},
\{N_{B_8}\}_{k_{B_8}}
 \}
 $\\
\noindent The variables are denoted by $X_1, X_2, Y_1$ and $Y_2$;\ The static names are denoted by $N_{A_1}$, $A_1$, $k_{B_1}$, $B_2$, $N_{A_2}$, $k_{A_2}$, $B_3$, $k_{A_3}$, $k_{B_4}$, $A_{4}$, $k_{B_5}$, $B_7$, $N_{B_7}$, $k_{A_6}$, $N_{B_8}$ and $k_{B_8}$.\\
\subsection{Results and Interpretation}
\noindent The results of analysis of the NSL protocol are summarized in Table \ref{NSLGrowth}.
\begin{table}[h]
\caption[]{\label{NSLGrowth} Conformity of NSL Protocol to Theorem \ref{PAT}}
   \centering
\begin{tabular}{|c|c|c|c|c|c|c|c|}
  \hline
  $\alpha$ &  Role & $R^-$ & $r^+$ & ${\mathcal{W}}'_{p,F}(\alpha, r^+)$& $\ulcorner \alpha \urcorner$ & $ F(\alpha,\partial[\overline{\alpha}] R^-)$& Theorem \ref{PAT}\\
  \hline
  $N_a^i$ & $A$& $\emptyset$ & $\{A.N_a^i\}_{k_b}$ & $\{A, B\}$ &  $\{A, B\}$& $ \top$& Fulfilled\\
  \hline
 $\forall X$ &  $A$ & $\{B.N_a^i\}_{k_a}.\{B.X\}_{k_a}$ & $A.B.\{X\}_{k_b}$ & $\{B\}$& $\ulcorner X \urcorner$ & $ \{A, B\}$& Fulfilled\\
  \hline
 \hline
 $\forall Y$ & $B$ & $\{A.Y\}_{k_b}$ & $\{B.Y\}_{k_a}.\{B.N_b^i\}_{k_a}$ & $\{A, B\}$& $\ulcorner Y \urcorner$ & $ \{A, B\}$& Fulfilled\\
 \hline
  $N_b^i$ & $B$ & $\{A.Y\}_{k_b}$ & $\{B.Y\}_{k_a}.\{B.N_b^i\}_{k_a}$ & $\{A, B\}$& $\{A, B\}$ & $ \{A, B\}$& Fulfilled\\
   \hline
\end{tabular}
\\\hspace{\linewidth}

\end{table}
\noindent We notice from Table \ref{NSLGrowth} that the NSL protocol respects the correctness criterion stated in Theorem \ref{PAT}, therefore it is correct with respect to the secrecy property.

%% file: ConfconclusionEK.tex
\section{\sc\textbf{Comparison with related works}}
Houmani et al. in ~\cite{Houmani1,Houmani3,Houmani8,Houmani5} defined universal functions to analyze protocols. Although they gave a comprehensive guideline to build these functions, only two functions had been concretely defined: DEK and DEKAN. That is due to the difficulty to find, and after to prove, that a function satisfies to the property of full-invariance by substitution. With a Witness-Function, there is no need to this property. A Witness-Function operates on derivative messages instead of messages of the generalized roles of a protocol which reduces variables' effect. It also offers two helpful bounds that do not depend on substitutions. This relaxes the conditions made on the associated interpretation functions and allows us to have more reliable functions  and therefore  to prove secrecy in a wider range of protocols. In addition, our method does not need any complicated context as do the rank-functions of Schneider in~\cite{Schneider4} that require the CSP formalism~\cite{Schneider2}. It does not need neither  a strong message-typing as required by Abadi in~\cite{AbadiTyping}. Our method joins though Houmani's method and Schneider's method in the way of seeing secrecy in a protocol through its growth. 
\section{\sc\textbf{Conclusion and future work}}

\noindent In this paper, we suggested a new semi-decidable procedure to analyze cryptographic protocols with the Witness-Functions based on selection inside the external protective key  for the property of secrecy. We experimented our approach on real protocols and we showed that the Witness-functions can even help to teach about flaws. They were successful to prove secrecy in others too.  As a future work, we intend to define more Witness-Functions based on selection inside other keys like the most internal key or a group of  keys together to enlarge the range of protocols that could be proved correct. We intend also to check our approach on protocols with non-empty theories~\cite{cortier9900,cortier1971,cortier9901}. In addition, we intend to take advantage of the bounds offered by Witness-Functions so that we can use exchanged messages in a protocol as encryption keys. We think also to extend our approach so that it can deal with other security properties like authentication and integrity. 